\documentclass{article}
\usepackage{t1enc}      % needed to have the edth and thorn operators

\usepackage{amsmath}
\usepackage{amssymb}
\usepackage{mathrsfs}  % needed to have the correct character for scri

\newcommand{\edth} {\mbox{\symbol{'360}}}

\newcommand{\ua}{\underline a \,}

\newcommand{\uA}{\underline A \,}

\newcommand{\bi}{\bf i} 
\newcommand{\bj}{\bf j} 
\newcommand{\bk}{\bf k}

\numberwithin{equation}{section}

\begin{document}
\bibliographystyle{unsrt}

\title{Total angular momentum from Dirac eigenspinors}

\author{ L\'aszl\'o B Szabados \\
Research Institute for Particle and Nuclear Physics \\
H-1525 Budapest 114, P. O. Box 49, Hungary \\
e-mail: lbszab@rmki.kfki.hu }
\maketitle

\begin{abstract}
The eigenvalue problem for Dirac operators, constructed from two 
connections on the spinor bundle over closed spacelike 2-surfaces, 
is investigated. A class of divergence free vector fields, built 
from the eigenspinors, are found, which, for the lowest eigenvalue, 
reproduce the rotation Killing vectors of metric spheres, and 
provide rotation BMS vector fields at future null infinity. This 
makes it possible to introduce a well defined, gauge invariant 
spatial angular momentum at null infinity, which reduces to the 
standard expression in stationary spacetimes. The general formula 
for the angular momentum flux carried away be the gravitational 
radiation is also derived. 

\end{abstract}

%%%%%%%%%%%%%%%%%%%%%%%%%%%%%%%%%%%%%%%%%%%%%%%%%%%%%%%%%%%%%%%%%%%%%%

\section{Introduction}
\label{sec-1}

Angular momentum is one of the basic conserved quantities in physics, 
and in general relativity there is, indeed, a well defined notion of 
total (ADM) angular momentum of an isolated system `measured' at {\em 
spatial infinity}. On the other hand, if we are interested in the 
angular momentum carried away by the gravitational radiation, then 
we should be able to define total angular momentum at future {\em null 
infinity}, too. Unfortunately, however, there is no generally accepted 
notion of angular momentum at null infinity of a radiative spacetime. 
(For a review of classical results see \cite{Win}, and for the recent 
ones see e.g. \cite{Sz01,Mo,CJK,Sz04,PRII,He}.) Another difficulty is 
that even if we have an ambiguity-free notion of angular momentum, it 
is not guaranteed that the angular momenta measured at {\em different} 
retarded times (i.e. when they are associated with {\em different} 
cuts of future null infinity) can be compared, and hence there is no 
unambiguous way of computing the angular momentum radiated away in a 
given time interval by the localized source (see e.g. 
\cite{Win,PRII}). 

The situation at the quasi-local level (i.e. when we are considering 
only subsystems of the whole universe, or, mathematically, we intend 
to associate physical quantities with closed spacelike 2-surfaces 
${\cal S}$) is even worse, because we do not have even the asymptotic 
(BMS) symmetries. The best that we can do is to try to systematically 
`quasi-localize' the canonical analysis of GR. Although this program 
is not yet completed, the first few steps towards the quasi-local 
canonical GR have been made \cite{Sz06}, yielding a class of well 
defined, 2+2 covariant, gauge invariant observables $O[N^a]$. These 
are based on divergence free vector fields $N^a$ that are tangent to 
the 2-surface ${\cal S}$. Unfortunately, however, without additional 
restriction on the vector fields these observables reflect properties 
of the 2-surface as a submanifold in the spacetime, rather than the 
properties of the gravitational `field' itself: $O[N^a]$ may be 
non-zero for 2-surfaces even in Minkowski spacetime. Thus to obtain 
physically interesting observables in the form of $O[N^a]$, new ideas 
are needed how to choose the vector fields $N^a$. 

On the other hand, and remarkably enough, the requirement of the 
finiteness of $O[N^a]$ at spatial infinity of asymptotically flat 
spacetimes already restricts the asymptotic structure of $N^a$ such 
that the corresponding observable will be just the familiar spatial 
(ADM) angular momentum. Similarly, at future null infinity of a {\em 
stationary} asymptotically flat spacetime $O[N^a]$ reproduces the 
standard angular momentum expression. Nevertheless, in radiative 
spacetimes at future null infinity $O[N^a]$ is still ambiguous. Thus 
to obtain a well-defined notion of angular momentum at future null 
infinity a more detailed prescription of the divergence-free vector 
fields should be given. 

A potentially viable construction might be based on the recent idea of 
approximate Killing vectors on topological 2-spheres \cite{CW}. These 
are the divergence-free vector fields that solve a variational problem, 
whose action is built form the norm of the Killing operator. Since 
these vector fields are divergence free by construction, they can be 
used in $O[N^a]$ to get well-defined, gauge invariant observables. 

Another promising approach of constructing such observables could be 
based on the spectral analysis of the Laplace, or of the Dirac 
operators on closed spacelike 2-surfaces. Clearly, the eigenvalues of 
these operators are gauge invariant, and they should reflect, maybe 
in a rather implicit way, the geometrical properties of the surface. 
(For example, the mass and angular momentum parameters can be recovered 
from the eigenvalues of the Laplacian on the event horizon of a 
Kerr--Newman black hole \cite{ECS}.) Indeed, in Riemannian geometry 
mathematicians already proved the existence of sharp lower bounds to 
the first eigenvalue of the Dirac operator in terms of the scalar 
curvature \cite{TFr,Hi86,Hi95,TF00} or the volume \cite{Ba92,TF00,FK}. 
Similar results exist for hypersurface Dirac operators when the lower 
bounds are given in terms of the curvature of the intrinsic geometry 
and the extrinsic curvature \cite{Zh}. 

In the present paper we also investigate the eigenvalue problem of the 
Dirac operators. However, instead of the eigenvalues, we concentrate 
on the (almost always) overlooked eigenspinors. A further difference 
between the former and the present investigations is that the base 
manifold on which the Dirac operators are defined is not simply a 
two-dimensional Riemannian manifold, but a spacelike 2-surface in a 
general Lorentzian spacetime. Moreover, the two Dirac operators that 
we study here are built from both the intrinsic and extrinsic 
geometrical objects of the surface. This yields a number of 
conceptual difficulties (e.g. no natural constant Hermitian metric 
exists on the spinor bundle), and hence we can define the eigenvalue 
problem only for the spacetime (rather than the 2-surface) Dirac 
spinors. Moreover, in lack of any natural Hermitian metric the reality 
of the eigenvalues is not guaranteed. Nevertheless, we show how the 
Nester--Witten integral can be used to give criteria for the reality 
of the eigenvalues. We construct three divergence-free vector fields 
from the eigenspinors. In particular, we show that on round spheres 
with radius $r$ two of these vector fields, built from the eigenspinors 
with the lowest eigenvalue of one of the two Dirac operators, are 
proportional to the $1/r$ times, and the third to the $1/r^2$ times 
the rotation Killing vectors of the metric sphere. 

Based on this observation we show that on the large spheres $u={\rm 
const}$, $r={\rm const}$ in a Bondi type coordinate system near the 
future null infinity in an asymptotically flat spacetime the 
divergence-free vector fields analogous to the first two above 
determine the rotation BMS vector fields tangent to the cut $u={\rm 
const}$ of ${\mathscr{I}}^+$. As we already mentioned, we expect that 
the gravitational energy-momentum and angular momentum be connected 
with the Hamiltonian formulation of Einstein's theory, i.e. these 
observables are expected to be the value of the correct Hamiltonian 
of the theory with appropriately chosen generators. 
Thus it seems natural to define the angular momentum at the future 
null infinity by the observable $O[N^a]$ in which the vector fields 
$N^a$ are chosen to be the divergence-free vector fields built from 
the eigenspinors with the lowest eigenvalue of the Dirac operator. 
This strategy gives two observables: the first can be interpreted as 
spatial angular momentum. This is gauge invariant, free of ambiguities, 
and trivially reduces to the standard expressions in the stationary 
and in the axi-symmetric spacetimes. Moreover, there is a natural way 
of comparing the angular momentum measured at different retarded times, 
and hence we can compute the angular momentum flux carried away by 
the gravitational radiation. 
The other observable is a non-negative expression of the magnetic part 
of the asymptotic shear, and it is vanishing precisely when the shear 
is purely electric. Since the shear is known to be purely electric in 
stationary spacetimes, this is a measure of dynamics of the 
gravitational `field'  near the future null infinity. Its significance 
is, however, not yet clear. Remarkably enough, though the two BMS 
vector fields are different, they define the same pair of observables. 

The organization of the paper follows the logic of the results above: 
in the second section we recall the basic notions, discuss the general 
aspects of the eigenvalue problem for the Dirac operator that is based 
on a Sen-type derivative operator, construct the divergence-free vector 
fields and discuss the reality properties of the eigenvalues. Section 
3 is devoted to the discussion of the special properties of another 
Dirac operator, which is a reduction of the previous one and is built 
only from the intrinsic geometry and the connection 1-form of the 
normal bundle of the 2-surface. The related non-existence result for 
the {\em constant} positive definite Hermitian scalar product in given 
in the appendix. Then, in section 4, we apply these ideas to round 
spheres, where we calculate the spectrum of the Dirac operators and 
construct the divergence-free vector fields explicitly. Section 5 is 
devoted to the analogous calculations on large spheres near the future 
null infinity in asymptotically flat spacetimes. We also calculate the 
corresponding angular momentum and angular momentum flux as well. 

The signature of $g_{ab}$ is $(+,-,-,-)$, the curvature and Ricci 
tensors and the curvature scalar are defined by $R^a{}_{bcd}X^b:=
-(\nabla_c\nabla_d-\nabla_d\nabla_c)X^a$, $R_{bd}:=R^a{}_{bad}$ and 
$R:=R_{ab}g^{ab}$, respectively. Then Einstein's equations take the 
form $G_{ab}=-8\pi GT_{ab}$, where $G$ is Newton's gravitational 
constant.

%%%%%%%%%%%%%%%%%%%%%%%%%%%%%%%%%%%%%%%%%%%%%%%%%%%%%%%%%%%%%%%%%

\section{The $\Delta_{AA'}$-Dirac operator}
\label{sec-2}

\subsection{Generalities}
\label{sub-2.1}

Let ${\mathbb S}^A({\cal S})$ denote the bundle of 2-component (i.e. 
Weyl) spinors over the closed, orientable spacelike 2-surface ${\cal 
S}$ in the space and time orientable spacetime, and we denote the 
complex conjugate bundle by 
$\bar{\mathbb S}^{A'}({\cal S})$. From the spacetime structure two 
metrics are inherited: the symplectic $\varepsilon_{AB}$ and the 
symmetric $\gamma_{AB}$. The former is just the spinor form of the 
spacetime metric, $g_{ab}=\varepsilon_{AB}\varepsilon_{A'B'}$, while 
the latter is built from the timelike and spacelike unit normals of 
${\cal S}$, $t^{AA'}$ and $v^{AA'}$, respectively, as $\gamma^A{}_B
:=2t^{AA'}v_{BA'}$. These normals define the projection $\Pi^a_b:=
\delta^a_b-t^at_b+v^av_b$ to the 2-surface, by means of which the 
restriction to ${\cal S}$ of the spacetime tangent bundle decomposes 
in a unique way to the $g_{ab}$-orthogonal direct sum of the tangent 
bundle $T{\cal S}$ and the normal bundle $N{\cal S}$. Though the 
actual normals $t^a$ and $v^a$ are not uniquely determined, both 
the projection $\Pi^a_b$ and the spinor $\gamma^A{}_B$ are well 
defined. This $\gamma^A{}_B$ defines a chirality on the spinor 
bundle ${\mathbb S}^A({\cal S})$. Thus the spinor bundle is endowed 
in a natural way by the symplectic metric and the chirality, and in 
this case the elements of $({\mathbb S}^A({\cal S}),\varepsilon_{AB},
\gamma^A{}_B)$ are called 2-surface spinors. 

On the spinor bundle, two connections can be introduced in a natural 
way, and the corresponding derivative operators will be denoted by 
$\delta_e$ and $\Delta_e$, respectively. The first is built from 
the intrinsic 2-metric and the connection 1-form $A_e:=\Pi^f_e(\nabla
_ft_a)v^a$ on the normal bundle of ${\cal S}$, while the other 
contains the spinor form $Q^A{}_{eB}$ of the extrinsic curvature 
tensor $Q^a{}_{eb}$ of ${\cal S}$ too, where $Q^A{}_{eB}=\frac{1}{2}
\Pi^f_e(\nabla_f\gamma^A{}_E)\gamma^E{}_B$. This $Q^A{}_{EE'B}$ has 
the reality property $Q^A{}_{AE'B}=\bar Q^{A'}{}_{A'BE'}$, 
expressing the hypersurface orthogonality of the two null normals of 
${\cal S}$ (or, in other words, the reality of the two individual 
convergences $\rho$ and $\rho'$ corresponding to the outgoing and 
incoming null normals, respectively). The action of the covariant 
derivative $\Delta_e$ is given explicitly by $\Delta_e\lambda^A=
\delta_e\lambda^A+Q^A{}_{eB}\lambda^B$. This is nothing but the 
projection to ${\cal S}$ of the spacetime Levi-Civita covariant 
derivative: $\Delta_a:=\Pi^b_a\nabla_b$. Both $\Delta_e$ and 
$\delta_e$ annihilate $\varepsilon_{AB}$, but $\gamma_{AB}$ is 
annihilated only by $\delta_e$. The curvatures $f^A{}_{Bcd}$ and 
$F^A{}_{Bcd}$ corresponding to $\delta_e$ and $\Delta_e$, 
respectively, are given by 

\begin{eqnarray}
f^A{}_{Bcd}\!\!\!\!&=\!\!\!\!&-\frac{\rm i}{4}f\gamma^A{}_B
 \varepsilon_{cd}, \label{eq:2.1.a}\\
F^A{}_{Bcd}\!\!\!\!&=\!\!\!\!&f^A{}_{Bcd}-\bigl(\delta_cQ^A{}_{dB}-
 \delta_dQ^A{}_{cB}+Q^A{}_{cE}Q^E{}_{dB}-Q^A{}_{dE}Q^E{}_{cB}\bigr); 
 \label{eq:2.1.b}
\end{eqnarray}
where $f:=f_{abcd}\frac{1}{2}(\varepsilon^{ab}-{\rm i}{}^\bot
\varepsilon^{ab})\varepsilon^{cd}=R-2{\rm i}\varepsilon^{ab}\delta_a
A_b$, the ``scalar curvature'' of the curvature $f^a{}_{bcd}$ of the 
derivative $\delta_e$ on the Lorentzian vector bundle over ${\cal 
S}$. Here $R$ is the scalar curvature of the intrinsic geometry of 
${\cal S}$, and $\varepsilon_{ab}$ and ${}^\bot\varepsilon_{ab}$ 
are the volume 2-forms on the tangent and normal 2-spaces of ${\cal 
S}$, respectively. (For the details see \cite{Sz94,Sz04}.) 

The $\Delta_{AA'}$-Dirac operator on ${\cal S}$ is defined in the 
decomposition of $\Delta_{A'A}\lambda_B$ in its unprimed indices 
to its anti-symmetric and $\gamma_{AB}$-trace-free symmetric parts: 
The former, $\Delta_{A'A}\lambda^A$, gives the Dirac operator, while 
the latter the 2-surface twistor operator of Penrose. (The $\gamma
_{AB}$ trace gives essentially the Dirac operator too: $\gamma^{AB}
\Delta_{A'A}\lambda_B=\bar\gamma_{A'}{}^{B'}\Delta_{B'B}\lambda^B$.) 
The square of the $\Delta_{AA'}$-Dirac operator is 

\begin{eqnarray}
\Delta_A{}^{A'}\Delta_{A'}{}^B\lambda_B=-\frac{1}{2}\Delta_e\Delta^e
 \lambda_A\!\!\!\!&-\!\!\!\!&\varepsilon^{A'B'}\varepsilon^{BC}Q
 _{e[ab]}\Delta^e\lambda_C-\nonumber \\
\!\!\!\!&-\!\!\!\!&\frac{1}{2}\varepsilon^{A'B'}\varepsilon^{BC}
 F_{CDAA'BB'}\lambda^D. \label{eq:2.2}
\end{eqnarray}
Taking into account (\ref{eq:2.1.a})-(\ref{eq:2.1.b}) and that the 
anti-symmetric part of the extrinsic curvature tensor is $2Q_{e[ab]}
=-(\varepsilon_{A'B'}Q_{AEE'B}+\varepsilon_{AB}\bar Q_{A'EE'B'})$, 
we find that 

\begin{eqnarray}
-2\Delta_A{}^{A'}\Delta_{A'}{}^B\lambda_B\!\!\!\!&=\!\!\!\!&\Delta_e
 \Delta^e\lambda_A+\frac{1}{4}f\lambda_A-2Q_{AeB}\Delta^e\lambda^B-
 \nonumber \\
-\varepsilon^{A'B'}\!\!\!\!\!\!&\Bigl(\!\!\!\!\!\!&\delta_{AA'}Q^B{}
 _{BB'C}-\delta_{BB'}Q^B{}_{AA'C}+ \nonumber \\
\!\!\!\!&+\!\!\!\!&Q^B{}_{AA'E}Q^E{}_{BB'C}-Q^B{}_{BB'E}Q^E
 {}_{AA'C}\Bigr)\lambda^C. \label{eq:2.3}
\end{eqnarray}
This equation is analogous to the Lichnerowicz identity \cite{L}: the 
square of the Dirac operator is expressed in terms of the Laplacian 
and the curvature, but here $\Delta_e\Delta^e$ is not the intrinsic 
Laplacian and the first derivative of the spinor field also appears 
on the right. 

An analogous analysis can be carried out with the $\delta_{AA'}$-Dirac 
operator too, but all the results can be recovered from those for the 
$\Delta_{AA'}$-Dirac operator by the formal substitution $Q^A{}_{eB}
=0$ too. In particular, the identity (\ref{eq:2.3}) reduces to 

\begin{equation}
-2\delta_A{}^{A'}\delta_{A'}{}^B\lambda_B=\delta_e\delta^e\lambda_A
+\frac{1}{4}f\lambda_A. \label{eq:2.4}
\end{equation}
Note that, strictly speaking, this is still not the Lichnerowicz 
identity, because $\delta_{AA'}$ is {\em not} only an intrinsic 
derivative operator: it contains extrinsic quantities in the form of 
$A_e$ as well, while the Dirac operator in the genuine Lichnerowicz 
identity is built exclusively from the intrinsic geometry of ${\cal 
S}$.

%%%%%%%%%%%%%%%%%%%%%%%%%%%%%%%%%%%%%%%%%%%%%%%%%%%%%%%%%%%%%%%%%%

\subsection{The eigenvalue problem for the 2-surface Dirac operators}
\label{sub-2.2}

If we had a naturally defined Hermitian metric $G_{AA'}$ on ${\mathbb 
S}^A({\cal S})$ by means of which the bundles ${\mathbb S}^A({\cal 
S})$ and $\bar{\mathbb S}^{A'}({\cal S})$ could be identified (i.e. 
the primed indices could be converted to unprimed ones), then the 
eigenvalue equation for the $\delta_{AA'}$-Dirac operator could be 
defined as 

\begin{equation}
{\rm i}G_A{}^{A'}\delta_{A'}{}^B\lambda_B=-\frac{\alpha}{\sqrt{2}}
\lambda_A. \label{eq:2.a}
\end{equation}
(The choice for the apparently ad hoc coefficient $-1/\sqrt{2}$ in 
front of the eigenvalue $\alpha$ yields the compatibility both with 
the subsequent more general analysis and the known standard results 
in special cases.) 
However, it is desirable that, in addition, such a Hermitian metric 
be compatible with the connection in the sense that $\delta_eG_{AA'}
=0$. Nevertheless, in the appendix we show that the existence of 
such a Hermitian metric is equivalent to the vanishing of the 
holonomy of $\delta_e$ on the normal bundle (and, in particular, its 
curvature, ${\rm Im}\,f$, must be zero). 
Therefore, on a general 2-surface in a general, curved spacetime we 
do not have any such natural Hermitian structure on ${\mathbb S}^A
({\cal S})$. Thus to motivate how the eigenvalue problem should be 
defined for the $\delta_{AA'}$ (or, more generally, for the $\Delta
_{AA'}$)-Dirac operator, let us consider the eigenvalue problem for 
the spacetime Dirac (rather than the Weyl) spinors, where the primed 
and unprimed indices are treated on equal footing. 

Recall that a Dirac spinor $\Psi^\alpha$ is a pair of Weyl spinors 
$\lambda^A$ and $\bar\mu^{A'}$, written them as a column vector 

\begin{equation}
\Psi^\alpha=\left(\begin{array}{cc}\lambda^A \\  
                          \bar\mu^{A'}\end{array}\right) 
\label{eq:2.5}
\end{equation}
and adopting the convention $\alpha=A\oplus{A'}$, $\beta=B\oplus{B'}$ 
etc. Its derivative $\Delta_e\Psi^\alpha$ is the column vector 
consisting of $\Delta_e\lambda^A$ and $\Delta_e\bar\mu^{A'}$. If 
Dirac's $\gamma$-`matrices' are denoted by $\gamma^\alpha_{e\beta}$, 
then one can consider the eigenvalue problem 

\begin{equation}
{\rm i}\gamma^\alpha_{e\beta}\Delta^e\Psi^\beta=\alpha\Psi^\alpha.  
\label{eq:2.6}
\end{equation}
Explicitly, with the representation 

\begin{equation}
\gamma^\alpha_{e\beta}=\sqrt{2}\left(\begin{array}{cc}
          0&\varepsilon_{E'B'}\delta^A_E \\
      \varepsilon_{EB}\delta^{A'}_{E'}&0 \end{array}\right) 
\label{eq:2.7}
\end{equation}
(see e.g. \cite{PRI}, pp 221), this is just the pair of equations 

\begin{equation}
{\rm i}\Delta_{A'}{}^A\lambda_A=-\frac{\alpha}{\sqrt2}\bar\mu_{A'}, 
\hskip 20pt
{\rm i}\Delta_A{}^{A'}\bar\mu_{A'}=-\frac{\alpha}{\sqrt2}\lambda_A.
\label{eq:2.8}
\end{equation}
By (\ref{eq:2.8}) $\Psi^\alpha=(\lambda^A,\bar\mu^{A'})$ (as a column 
vector) is a Dirac eigenspinor with eigenvalue $\alpha$ precisely 
when $(\lambda^A,-\bar\mu^{A'})$ is a Dirac eigenspinor with 
eigenvalue $-\alpha$. In the language of Dirac spinors this is 
formulated in terms of the chirality, represented by the so-called 
`$\gamma_5$-matrix', denoted here by 

\begin{equation}
\eta^\alpha{}_\beta:=\frac{1}{4!}\varepsilon^{abcd}\gamma^\alpha
_{a\mu}\gamma^\mu_{b\nu}\gamma^\nu_{c\rho}\gamma^\rho_{d\beta}=
{\rm i}\left(\begin{array}{cc} \delta^A_B&0\\ 
        0&-\delta^{A'}_{B'}\end{array}\right) 
\label{eq:2.9}
\end{equation}
(see appendix II. of \cite{PRII}). Since this is anti-commuting with 
$\gamma^\alpha_{e\beta}$, from (\ref{eq:2.6}) we obtain that ${\rm i}
\gamma^\alpha_{e\mu}\Delta^e(\eta^\mu{}_\beta\Psi^\beta)=-\alpha(\eta
^\alpha{}_\beta\Psi^\beta)$. Thus if $\Psi^\alpha$ is a Dirac 
eigenspinor with eigenvalue $\alpha$, then, in fact, $\eta^\alpha{}
_\beta\Psi^\beta$ is a Dirac eigenspinor with eigenvalue $-\alpha$. 

On the other hand, the Dirac eigenspinors with definite chirality 
belong to the kernel of the Dirac operator. Indeed, Dirac spinors 
with definite chirality have the structure either $(\lambda^A,0)$ or 
$(0,\bar\mu^{A'})$, which, by (\ref{eq:2.8}), yield that $\Delta
_{A'A}\lambda^A=0$ or $\Delta_{AA'}\bar\mu^{A'}=0$, respectively. 
Therefore, this notion of chirality cannot be used to decompose the 
space of the eigenspinors with given eigenvalue. Its role is simply 
to take a Dirac eigenspinor with eigenvalue $\alpha$ to a Dirac 
eigenspinor with eigenvalue $-\alpha$. 

The equations of (\ref{eq:2.8}) imply that 

\begin{equation}
-\Delta_A{}^{A'}\Delta_{A'}{}^B\lambda_B=\frac{1}{2}\alpha^2\lambda_A,
\hskip 20pt
-\Delta_{A'}{}^A\Delta_A{}^{B'}\bar\mu_{B'}=\frac{1}{2}\alpha^2\bar
 \mu_{A'}. \label{eq:2.10}
\end{equation}
Thus if $\Psi^\alpha=(\lambda^A,\bar\mu^{A'})$ is a Dirac eigenspinor 
with eigenvalue $\alpha$, then its Weyl spinor parts $\lambda_A$ and 
$\bar\mu_{A'}$ are eigenspinors of the {\em second order} operator 
$-2\Delta_A{}^{A'}\Delta_{A'}{}^B$ and $-2\Delta_{A'}{}^A\Delta_A{}
^{B'}$, respectively, with the {\em same} eigenvalue $\alpha^2$. 
Conversely, if $\lambda_A$ is a Weyl eigenspinor of $-2\Delta_A{}^{A'}
\Delta_{A'}{}^B$ with non-zero eigenvalue $\alpha^2$, then $\Psi
^\alpha_\pm$, built from $\lambda^A$ and $\bar\mu^{A'}:=\mp(\sqrt{2}/
\alpha){\rm i}\Delta^{A'A}\lambda_A$, are Dirac eigenspinors with 
eigenvalue $\pm\alpha$, respectively, for which $\eta^\alpha{}_\beta
\Psi^\beta_\pm={\rm i}\Psi^\alpha_\mp$.  
Therefore, {\em there is a natural isomorphism between the space 
${\bf W}_{\alpha^2}$ of the Weyl eigenspinors of $-2\Delta_A{}
^{A'}\Delta_{A'}{}^B$ with eigenvalue $\alpha^2$ and the direct sum 
${\bf D}_{\alpha}\oplus{\bf D}_{-\alpha}$, where ${\bf D}_{\alpha}$ 
is the space of the Dirac eigenspinors of $\Delta_e$ with eigenvalue 
$\alpha$. The chirality operator $\eta^\alpha{}_\beta$ maps ${\bf D}
_{\pm\alpha}$ to ${\bf D}_{\mp\alpha}$.} 
For zero eigenvalue, $\alpha=0$, the Weyl eigenspinors define the 
kernel of the $\Delta_{AA'}$-Dirac operator, which, apart from 
exceptional 2-surfaces, is empty (see e.g. \cite{Sz94,Sz01}). Thus 
for generic 2-surfaces, the eigenvalues of the $\Delta_{AA'}$-Dirac 
operator are nonzero. 

Finally, let us translate the general equations into the language of 
the GHP formalism \cite{GHP,PRI}. Thus let us fix a normalized spinor 
dyad $\{o^A,\iota^A\}$ adapted to the 2-surface ${\cal S}$ (i.e. the 
complex null vectors $m^a:=o^A\bar\iota^{A'}$ and $\bar m^a:=\iota^A
\bar o^{A'}$ are tangent to ${\cal S}$). Then the GHP form of the 
$\Delta_{AA'}$-Dirac operator and the square of the $\Delta
_{AA'}$-Dirac operator are 

\begin{eqnarray}
\bar o^{A'}\Delta_{A'}{}^A\lambda_A\!\!\!\!&=\!\!\!\!&{\edth}'
 \lambda_0+\rho\lambda_1, \label{eq:2.11.a} \\
-\bar\iota^{A'}\Delta_{A'}{}^A\lambda_A\!\!\!\!&=\!\!\!\!&{\edth}
 \lambda_1+\rho'\lambda_0, \label{eq:2.11.b} \\
o^A\Delta_A{}^{A'}\Delta_{A'}{}^B\lambda_B\!\!\!\!&=\!\!\!\!&{\edth}
 {\edth}'\lambda_0+\bigl({\edth}\rho\bigr)\lambda_1-\rho\rho'
 \lambda_0, \label{eq:2.12.a} \\
\iota^A\Delta_A{}^{A'}\Delta_{A'}{}^B\lambda_B\!\!\!\!&=\!\!\!\!&
 {\edth}'{\edth}\lambda_1+\bigl({\edth}'\rho'\bigr)\lambda_0-\rho
 \rho'\lambda_1. \label{eq:2.12.b} 
\end{eqnarray}
Here we defined the spinor components by the conventions $\lambda_0:=
\lambda_Ao^A$ and $\lambda_1:=\lambda_A\iota^A$; ${\edth}$ and 
${\edth}'$ are the standard edth operators and $\rho$ and $\rho'$ are 
the standard GHP convergences corresponding to the outgoing and incoming 
null normals $o^A\bar o^{A'}$ and $\iota^A\bar\iota^{A'}$ of ${\cal S}$, 
respectively. 
The dimension of the kernel of the edth operators depends on the genus 
of the 2-surface ${\cal S}$ \cite{FSz}: let $p$ be any real number. 
Then on topological 2-spheres $\dim\ker{\edth}_{(p,p+n)}=\dim\ker
{\edth}'_{(p+n,p)}=0$ for any $n\in{\mathbb N}$, while $\dim\ker
{\edth}_{(p+n,p)}=\dim\ker{\edth}'_{(p,p+n)}=1+n$ for any $n=0,1,2,
...$. 
On tori $\dim\ker{\edth}_{(p,p+n)}=\dim\ker{\edth}'_{(p,p+n)}=1$ for 
any integer $n$. 
On surfaces with genus $g\geq2$ one has $\dim\ker{\edth}_{(p,p+n)}=
\dim\ker{\edth}'_{(p+n,p)}=0$ if $-n\in{\mathbb N}$, it is 1 if $n=0$, 
it is $g-1$ if $n=1$, it is $g$ if $n=2$ and it is $(n-1)(g-1)$ if 
$n>2$.

\subsection{Special vector fields built from the eigenspinors}
\label{sub-2.3}

The Weyl eigenspinors define several $\Delta_e$-divergence-free complex 
Lorentzian vector fields on ${\cal S}$. In fact, let $\lambda^A$ and 
$\bar\mu^{A'}$ be the unprimed and primed Weyl spinor parts of a Dirac 
eigenspinor $\Psi^\alpha$, respectively. Then contracting the first 
equation of (\ref{eq:2.8}) with $\bar\mu^{A'}$ and the second equation 
of (\ref{eq:2.8}) with $\lambda^A$, and adding them together we obtain 

\begin{equation}
\Delta_{AA'}(\lambda^A\bar\mu^{A'})=0; \label{eq:2.13}
\end{equation}
i.e. {\em $K^a:=\lambda^A\bar\mu^{A'}$ is a $\Delta_e$-divergence-free 
complex vector field on ${\cal S}$.} Similarly, the contraction of the 
first equation of (\ref{eq:2.8}) with $\bar\lambda^{A'}$ gives $\bar
\lambda^{A'}\Delta_{A'A}\lambda^A=-\frac{\rm i}{\sqrt2}\alpha\bar\Phi$, 
and the contraction of the second equation of (\ref{eq:2.8}) with $\mu
^A$ gives $\mu^A\Delta_{AA'}\bar\mu^{A'}=\frac{\rm i}{\sqrt2}\alpha
\Phi$, where $\Phi:=\lambda^A\mu_A$. These imply that 

\begin{equation*}
\Delta_{AA'}\bigl(\lambda^A\bar\lambda^{A'}\bigr)=\frac{\rm i}
{\sqrt2}\bigl(\bar\alpha\Phi-\alpha\bar\Phi\bigr),  \hskip 20pt
\Delta_{AA'}\bigl(\mu^A\bar\mu^{A'}\bigr)=\frac{\rm i}{\sqrt2}
 \bigl(\alpha\Phi-\bar\alpha\bar\Phi\bigr); 
\end{equation*}
and hence, for any $a,b\in{\mathbb C}$, that 

\begin{equation}
\Delta_{AA'}\Bigl(a\lambda^A\bar\lambda^{A'}+b\mu^A\bar\mu^{A'}\Bigr)
=\frac{\rm i}{\sqrt2}\Bigl(\bigl(a\bar\alpha+b\alpha\bigr)\Phi-\bigl(
a\alpha+b\bar\alpha\bigr)\bar\Phi\Bigr). \label{eq:2.14} 
\end{equation}
Therefore, {\em the real vector field $Z_\pm^a:=\lambda^A\bar\lambda
^{A'}\pm\mu^A\bar\mu^{A'}$ is $\Delta_e$-divergence free for purely 
imaginary/real eigenvalue $\alpha$}. 

These vector fields can be recovered as special cases of $V_e:=\Phi
_\alpha\gamma^\alpha_{e\beta}\Psi^\beta$, built from the Dirac 
eigenspinors $\Phi^\alpha$ and $\Psi^\alpha$. In fact, if 

\begin{equation*}
{\rm i}\gamma^\alpha_{e\beta}\Delta^e\Psi^\beta=\alpha\Psi^\alpha, 
\hskip 14pt 
{\rm i}\gamma^\alpha_{e\beta}\Delta^e\Phi^\beta=\beta\Phi^\alpha, 
\end{equation*}
then ${\rm i}\Delta^e(\Phi_\alpha\gamma^\alpha_{e\beta}\Psi^\beta)=
(\alpha-\beta)\Phi_\alpha\Psi^\alpha$. Therefore, if $\Phi^\alpha=
(\phi^A,\bar\omega^{A'})$ and $\Psi^\alpha=(\lambda^A,\bar\mu^{A'})$ 
are eigenspinors with the same eigenvalue, say $\alpha$, then the 
vector field $V_e:=\Phi_\alpha\gamma^\alpha_{e\beta}\Psi^\beta=-
\sqrt{2}(\lambda_E\bar\omega_{E'}+\phi_E\bar\mu_{E'})$ is $\Delta
_e$-divergence free. In particular, (1) if $\Phi_\alpha=\Psi_\alpha$, 
then $V^e=-2\sqrt{2}\lambda^E\bar\mu^{E'}=-2\sqrt{2}K^e$; (2) if 
$\Phi^+_\alpha=\Psi_\alpha$ (i.e. $\omega_A=\lambda_A$ and $\phi_A=
\mu_A$), and hence $\Psi^\alpha$ is an eigenspinor with the eigenvalue 
$-\bar\alpha$ too and $\alpha$ is imaginary, then $V^e=-\sqrt{2}
(\lambda^E\bar\lambda^{E'}+\mu^E\bar\mu^{E'})=-\sqrt{2}Z^e_+$; 
(3) if $\Phi^+_\alpha\eta^\alpha{}_\beta=\Psi_\beta$ (i.e. ${\rm i}
\omega_A=\lambda_A$ and ${\rm i}\phi_A=\mu_A$), and hence $\Psi^\alpha$ 
is an eigenspinor with the eigenvalue $\bar\alpha$ too and $\alpha$ 
is real, then $V^e=-{\rm i}\sqrt{2}(\lambda^E\bar\lambda^{E'}-\mu^E
\bar\mu^{E'})=-{\rm i}\sqrt{2}Z^e_-$. 

Another class of special vector fields can be constructed purely from 
the unprimed Weyl spinor parts of one or two Dirac eigenspinors and 
their derivatives. Next we consider these. Contracting the identity 
(\ref{eq:2.3}) with an arbitrary spinor field $\phi^A$, using the 
relationship between the two derivative operators $\Delta_{EE'}$ and 
$\delta_{EE'}$ and the reality property $Q^A{}_{AE'B}=\bar Q^{A'}{}
_{A'BE'}$, a straightforward calculation gives 

\begin{eqnarray*}
-2\phi^A\Delta_A{}^{A'}\Delta_{A'}{}^B\lambda_B\!\!\!\!&=\!\!\!\!&
 \phi^A\Delta_e\Delta^e\lambda_A+\frac{1}{4}f\phi^A\lambda_A-2Q_{AeB}
 \phi^A\Delta^e\lambda^B-\\
-\bigl(\Delta^eQ_{AeB}\bigr)\phi^A\lambda^B\!\!\!\!&+\!\!\!\!&2
 \bigl(\Delta_A{}^{A'}Q^D{}_{DA'B}\bigr)\lambda^A\phi^B+\frac{1}{2}
 \lambda^A\phi_AQ_{BeD}Q^{BeD}.
\end{eqnarray*}
Interchanging the spinor fields $\lambda_A$ and $\phi_A$, adding 
the resulting formula to the old one and using $\phi^A\Delta_e
\Delta^e\lambda_A$ $=\Delta_e(\phi^A\Delta^e\lambda_A)-(\Delta_e
\phi^A)(\Delta^e\lambda_A)$, we obtain the geometric identity 

\begin{eqnarray}
\delta_e\Bigl(\phi^A\delta^e\lambda_A+\lambda^A\delta^e\phi_A
 \Bigr)\!\!\!\!&=\!\!\!\!&\Delta_e\Bigl(\phi^A\Delta^e\lambda_A+
 \lambda^A\Delta^e\phi_A-2 Q_A{}^e{}_B\lambda^A\phi^B\Bigr)=
 \nonumber \\
=\!\!\!\!&-\!\!\!\!&2\Bigl(\phi^A\Delta_A{}^{A'}\Delta_{A'}{}^B
 \lambda_B+\lambda^A\Delta_A{}^{A'}\Delta_{A'}{}^B\phi_B\Bigr)-
 \nonumber \\
\!\!\!\!&-\!\!\!\!&2\bigl(\Delta_A{}^{A'}Q^D{}_{DA'B}\bigr)\Bigl(
 \lambda^A\phi^B+\phi^A\lambda^B\Bigr). \label{eq:2.15}
\end{eqnarray}
If the last term vanishes, e.g. when $\Delta_{(A}{}^{A'}Q_{B)A'D}{}
^D=0$ holds, and if there is a function $\alpha:{\cal S}\rightarrow
{\mathbb C}$ such that 

\begin{equation}
-\Delta_A{}^{A'}\Delta_{A'}{}^B\lambda_B=\frac{1}{2}\alpha^2\lambda_A, 
\hskip 20pt
-\Delta_A{}^{A'}\Delta_{A'}{}^B\phi_B=\frac{1}{2}\alpha^2\phi_A, 
\label{eq:2.16}
\end{equation}
then the vector field 

\begin{equation}
\xi^e:=\phi^A\delta^e\lambda_A+\lambda^A\delta^e\phi_A 
\label{eq:2.17}
\end{equation}
is $\delta_e$ (and, in fact, $\Delta_e$)-divergence free. Therefore, 
in particular when $\lambda_A$ and $\phi_A$ are Weyl eigenspinors 
with the same eigenvalue (e.g. if $\phi_A=\lambda_A$), then in the 
special case $\Delta_{(A}{}^{A'}Q_{B)A'D}{}^D=0$ the (in general 
complex) tangent vector field $\xi^e$ is divergence free on ${\cal 
S}$ (both with respect to $\delta_e$ and $\Delta_e$). Since in the 
GHP formalism $\Delta_{(A}{}^{A'}Q_{B)A'D}{}^D=({\edth}'\rho')o_Ao_B-
({\edth}\rho)\iota_A\iota_B$, the vanishing of this term is equivalent 
to ${\edth}\rho=0$, ${\edth}'\rho=0$, ${\edth}\rho'=0$ and ${\edth}'
\rho'=0$; i.e. the convergences are constant on ${\cal S}$. 

$\xi^e$ can be generalized to be $\Phi^\alpha\delta^e\Psi_\alpha+
\Psi^\alpha\delta^e\Phi_\alpha$, which is $\delta_e$-divergence free 
if $\Phi^\alpha$ and $\Psi^\alpha$ are Dirac eigenspinors with the 
same eigenvalue and $\Delta_{(A}{}^{A'}Q_{B)A'D}{}^D$ $=0$.

\subsection{The reality of the eigenvalues}
\label{sub-2.4}

The $\Delta_e$-divergence of $Z^a_-$ vanishes if it is built from 
eigenspinors with {\em real} eigenvalue. Thus we should find some 
criteria of the reality of the eigenvalues. The usual proof of the 
reality of the eigenvalues is based on the existence of a positive 
definite Hermitian metric compatible with the connection underlying 
the Dirac operator. However, in the light of the non-existence 
result even for the $\delta_e$-Dirac operator (see the appendix), 
the condition of this reality should be searched for following a 
different strategy. 
First, one might be tempted to define the eigenvalue problem 
(\ref{eq:2.8}) with the additional requirement that $\mu_A=c\lambda
_A$ for some complex constant $c$. (In the language of Dirac spinors 
a spinor $\Psi^\alpha$ with $\mu^A=\lambda^A$ is called a Majorana 
spinor.) However, by (\ref{eq:2.8}) this would imply that $\vert c
\vert=1$ and $\bar\alpha=-\alpha$, and hence that all the eigenvalues 
of $-\Delta_A{}^{A'}\Delta_{A'}{}^B$ would be non-positive. We will 
see that this cannot be the case: for round spheres the eigenvalues 
of $-\Delta_A{}^{A'}\Delta_{A'}{}^B$, maybe apart from finitely many 
of them, are all {\em positive}. 

Another strategy is to introduce a {\em global} Hermitian scalar 
product directly on the space $C^\infty({\cal S},{\mathbb S}^A)$ of 
all smooth spinor fields on ${\cal S}$, without trying to link this 
with any {\em pointwise} Hermitian scalar product $G_{AA'}$. This 
is based on the integral of the Nester--Witten 2-form built from 
the Weyl spinors \cite{HT}. For any pair $(\lambda_A,\omega_A)$ of 
spinor fields it can be rewritten in the form \cite{Sz94b}

\begin{equation}
H\bigl[\lambda_A,\bar\omega_{A'}\bigr]=\frac{1}{4\pi G}\oint_{\cal 
S}\bar\gamma^{A'B'}\bar\omega_{A'}\Delta_{B'}{}^B\lambda_B {\rm d}
{\cal S}. \label{eq:2.18}
\end{equation}
It is a straightforward calculation to show that it is Hermitian in 
the sense that $H[\lambda_A,\bar\omega_{A'}]=\overline{H[\omega_A,
\bar\lambda_{A'}]}$, where overline denotes complex conjugation. In 
particular, $H[\lambda_A,\bar\lambda_{A'}]$ is always real, but in 
general it may be negative or zero even for non-vanishing spinor 
fields. Then with the substitution $\bar\omega_{A'}:=\Delta_{A'}{}
^B\pi_B$ we obtain 

\begin{eqnarray}
\overline{H\bigl[\Delta_A{}^{B'}\bar\pi_{B'},\bar\lambda_{A'}\bigr]}
 \!\!\!\!&=\!\!\!\!&H\bigl[\lambda_A,\Delta_{A'}{}^B\pi_B\bigr]= 
 \frac{1}{4\pi G}\oint_{\cal S}\bar\gamma^{A'B'}\bigl(\Delta_{A'}{}
 ^A\pi_A\bigr)\bigl(\Delta_{B'}{}^B\lambda_B\bigr){\rm d}{\cal S} 
 \nonumber \\
\!\!\!\!&=\!\!\!\!&H\bigl[\pi_A,\Delta_{A'}{}^B\lambda_B\bigr]; 
\label{eq:2.19}
\end{eqnarray}
i.e. the $\Delta_{AA'}$-Dirac operator is compatible with the 
Hermitian scalar product. This implies that 

\begin{equation}
H\bigl[\Delta_A{}^{B'}\Delta_{B'}{}^B\sigma_B,\bar\lambda_{A'}\bigr]
=H\bigl[\sigma_A,\Delta_{A'}{}^B\Delta_B{}^{B'}\bar\lambda_{B'}\bigr]
\label{eq:2.20}
\end{equation}
for any pair $(\sigma_A,\lambda_A)$ of spinor fields; i.e. {\em 
$\Delta_A{}^{A'}\Delta_{A'}{}^B$ is formally self-adjoint with 
respect to $H$}. 

It might be worth noting that originally the Nester--Witten integral 
was introduced as a Hermitian quadratic form on the space of Dirac 
spinor fields \cite{Wi,Ne}, and, with the representation $\Psi^\alpha
=(\lambda^A,\bar\mu^{A'})$, this can be written as $H[\lambda_A,\bar
\lambda_{A'}]-H[\mu_A,\bar\mu_{A'}]$. A natural extension of 
(\ref{eq:2.18}) as a Hermitian bilinear form to Dirac spinors (and of 
the original Nester--Witten integral too) is the integral of 

\begin{equation*}
\frac{\rm i}{\sqrt{2}}\Phi_\alpha\gamma^\alpha_{e\beta}\varepsilon
^{ef}\Delta_f\Psi^\beta=\Bigl(\bar\gamma^{A'B'}\bar\omega_{A'}\Delta
_{B'}{}^B\lambda_B-\gamma^{AB}\phi_A\Delta_B{}^{B'}\bar\mu_{B'}\Bigr)
\end{equation*}
for any pair of Dirac spinors $\Psi^\alpha=(\lambda^A,\bar\mu^{A'})$ 
and $\Phi^\alpha=(\phi^A,\bar\omega^{A'})$. In fact, 

\begin{eqnarray}
\bigl(\Psi^\alpha,\Phi^\alpha\bigr)\mapsto H\bigl[\Psi_\alpha,\Phi^+
 _\alpha\bigr]:\!\!\!\!\!&=\!\!\!\!\!&\frac{1}{4\pi G}\oint_{\cal S}
 \frac{\rm i}{\sqrt{2}}\Phi^+_\alpha\gamma^\alpha_{e\beta}\varepsilon
 ^{ef}\Delta_f\Psi^\beta{\rm d}{\cal S}= \nonumber \\
\!\!\!\!\!&=\!\!\!\!\!&H\bigl[\lambda_A,\bar\phi_{A'}\bigr]-H\bigl[
 \omega_A,\bar\mu_{A'}\bigr] \nonumber
\end{eqnarray}
is Hermitian, and reduces to (\ref{eq:2.18}) if at least one of $\Psi
^\alpha$ and $\Phi^\alpha$ has definite $\eta^\alpha{}_\beta$-chirality 
(e.g. when $\omega^A=0$ or $\mu^A=0$). Interestingly enough, if the 
Dirac spinors have definite, but opposite $\eta^\alpha{}
_\beta$-chirality, e.g. when $\phi^A=0$ and $\mu^A=0$, then they are 
orthogonal to each other: $H[\Psi_\alpha,\Phi^+_\alpha]=0$. 

Returning to the characterization of the reality of the eigenvalues 
of the Dirac operators, suppose that $\lambda_A$ satisfies 
(\ref{eq:2.10}). Then (\ref{eq:2.20}) implies that $\alpha^2H[\lambda
_A,\bar\lambda_{A'}]=\bar\alpha^2H[\lambda_A,\bar\lambda_{A'}]$, i.e. 
{\em $\alpha$ is real or purely imaginary provided $H[\lambda_A,$ $\bar
\lambda_{A'}]\not=0$}. The reality of $\alpha$ can be characterized 
by the $H$-norm of the eigenspinors. Indeed, for $\lambda_A$ and $\mu
_A$ satisfying (\ref{eq:2.8}) with non-zero $\alpha$, (\ref{eq:2.18}) 
gives 

\begin{equation*}
\bar\alpha H\bigl[\lambda_A,\bar\lambda_{A'}\bigr]=
-\alpha H\bigl[\mu_A,\bar\mu_{A'}\bigr]. 
\end{equation*}
This implies that $H[\lambda_A,\bar\lambda_{A'}]=\pm H[\mu_A,\bar\mu
_{A'}]$, and, for $H[\lambda_A,\bar\lambda_{A'}]\not=0$, we obtain 
that {\em $\alpha$ is real iff $H[\lambda_A,\bar\lambda_{A'}]=- 
H[\mu_A,\bar\mu_{A'}]$, and $\alpha$ is purely imaginary iff 
$H[\lambda_A,\bar\lambda_{A'}]=H[\mu_A,\bar\mu_{A'}]$}. In subsection 
\ref{sub-4.1} we give examples both for real and purely imaginary 
eigenvalues. $\alpha$ can be a more general complex number only if 
$H[\lambda_A,\bar\lambda_{A'}]=0$, i.e. when $\oint_{\cal S}\gamma
^{AB}\lambda_A\mu_B{\rm d}{\cal S}=0$. 
If $\sigma_A$ is an eigenspinor of $-2\Delta_A{}^{B'}
\Delta_{B'}{}^B$ with eigenvalue $\beta^2$, then (\ref{eq:2.20}) 
gives that $(\beta^2-\bar\alpha^2)H[\sigma_A,\bar\lambda_{A'}]=0$; 
i.e. {\em the eigenspinors of $-2\Delta_A{}^{B'}\Delta_{B'}{}^B$ 
with eigenvalues $\alpha^2$ and $\beta^2$, satisfying $\bar\alpha
\not=\pm\beta$, are orthogonal to each other with respect to $H$}.

%%%%%%%%%%%%%%%%%%%%%%%%%%%%%%%%%%%%%%%%%%%%%%%%%%%%%%%%%%%%%%%%%%%%%%%

\section{The $\delta_{AA'}$--Dirac operator}
\label{sec-3}

Since ${\cal S}$ is even dimensional, there is a notion of chirality 
in the space of surface spinors too, which chirality remains intact 
even if the gauge group $SO(2)$ is enlarged to $SO(2)\times SO(1,1)$ 
by including the boost gauge transformations in the normal bundle of 
${\cal S}$. This chirality is represented by the spinor $\gamma^A{}
_B$ (thus we call it the `$\gamma$-chirality'), and is preserved by 
$\delta_e$ but {\em not} by $\Delta_e$. (For the details see 
\cite{Sz94,Sz04}.) Thus it seems useful to discuss the consequences 
of the existence of the $\gamma$-chirality in the case of the $\delta
_{AA'}$--Dirac operator. 

The $\delta_{AA'}$-Dirac operator can be obtained from the $\Delta
_{AA'}$--Dirac operator with the formal substitution $Q^A{}_{eB}=0$; 
and in this case the eigenvalue problem (\ref{eq:2.8}) reduces to 

\begin{equation}
{\rm i}\delta_{A'}{}^A\lambda_A=-\frac{\alpha}{\sqrt2}\bar\mu_{A'}, 
\hskip 20pt
{\rm i}\delta_A{}^{A'}\bar\mu_{A'}=-\frac{\alpha}{\sqrt2}\lambda_A.
\label{eq:3.1}
\end{equation}
It is easy to show (using e.g. the second order equation $-2\delta_A
{}^{A'}\delta_{A'}{}^B\lambda_B=\alpha^2\lambda_A$) that this notion 
of eigenspinors coincides with that defined by (\ref{eq:2.a}) if a 
constant Hermitian metric $G_{AA'}$ exists on the spinor bundle. In 
this case $\bar\mu_{A'}=G_{A'A}\lambda^A$. 

Recalling that $\delta_e\gamma^A{}_B=0$, $\gamma^A{}_B\gamma^B{}_C=
\delta^A_C$ and that $\gamma^A{}_B\bar\gamma^{A'}{}_{B'}$ acts on 
vectors tangent to ${\cal S}$ as $-\delta^a_b$, it is easy to verify 
that 

\begin{equation}
\bar\gamma_{A'}{}^{B'}\bigl(\delta_{B'}{}^B\lambda_B\bigr)=\delta_{A'}
{}^A\bigl(\gamma_A{}^B\lambda_B\bigr), 
\hskip 15pt
\gamma_A{}^B\bigl(\delta_B{}^{B'}\delta_{B'}{}^C\lambda_B\bigr)=
\delta_A{}^{A'}\delta_{A'}{}^B\bigl(\gamma_B{}^C\lambda_C\bigr). 
\label{eq:3.2}
\end{equation}
Thus the $\delta_{AA'}$-Dirac operator commutes with the action of the 
$\gamma$-spinor as a base point preserving bundle map. In particular, 
if $\lambda_A$ is a Weyl eigenspinor of $-2\delta_A{}^{A'}\delta_{A'}
{}^B$ with eigenvalue $\alpha^2$, then $\gamma_A{}^B\lambda_B$ is also 
a Weyl eigenspinor with the {\em same} eigenvalue. This implies that 
both $\lambda_A\pm\gamma_A{}^B\lambda_B$ are Weyl eigenspinors with 
the same eigenvalue $\alpha^2$, but they have {\em definite 
$\gamma$-chirality}. 
In the GHP spinor dyad $\{o^A,\iota^A\}$ the right/left handed Weyl 
eigenspinors of $-2\delta_A{}^{A'}\delta_{A'}{}^B$ have the structure 
$-\lambda_0\iota_A$ and $\lambda_1o_A$, respectively, where, by 
(\ref{eq:2.12.a})-(\ref{eq:2.12.b}), the spinor components satisfy 

\begin{equation}
-2{\edth}{\edth}'\lambda_0=\alpha^2\lambda_0, \hskip 20pt
-2{\edth}'{\edth}\lambda_1=\alpha^2\lambda_1. \label{eq:3.3} 
\end{equation}
Thus the $\gamma$-chirality can be used to decompose the space 
of the eigenspinors further. Since, however, $\gamma^A{}_B$ is 
annihilated only by $\delta_e$ but {\em not} by $\Delta_e$ in 
general, this decomposition is possible only for the $\delta
_{AA'}$-Dirac operators. 
Using the list of the dimension of the kernel of the edth operators, 
by (\ref{eq:3.3}) we can determine the number of the (e.g. right 
handed) eigenspinors of $-2\delta_A{}^{A'}\delta_{A'}{}^B$ with zero 
eigenvalue: on topological 2-spheres there are no such eigenspinors, 
on tori there are two, while on higher genus ($g>1$) surfaces there 
are $2(g-1)$ ones. 

Clearly, there is a natural one-to-one correspondence between the 
eigenspinors $\lambda_A$ of $-2\delta_A{}^{A'}\delta_{A'}{}^B$ with 
eigenvalue $\alpha^2$ and the eigenspinors $(\lambda_A,\bar\mu_{A'})$ 
of (\ref{eq:3.1}). Moreover, if $\lambda_A$ has definite chirality, 
e.g. if $\gamma_A{}^B\lambda_B=\pm\lambda_A$, then $\mu_A$ has the 
same definite chirality: $\gamma_A{}^B\mu_B=\pm\mu_A$. Thus the 
natural one-to-one correspondence above preserves the 
$\gamma$-chirality as well, and the space of the eigenspinors splits 
in a natural way to the direct sum of the spaces of the right/left 
handed eigenspinors. 

By the analysis of subsection \ref{sub-2.3}, the Weyl eigenspinors 
define a collection of $\delta_e$-divergence-free complex vector 
fields on ${\cal S}$. Indeed, taking into account that $\delta_e$ 
commutes with the projection $\Pi^a_b$, the complex vector field 
$k^a:=\Pi^a_b\lambda^B\bar\mu^{B'}$, tangent to ${\cal S}$, is 
$\delta_e$-divergence free on ${\cal S}$. Similarly, $z^a_\pm:=\Pi
^a_b(\lambda^B\bar\lambda^{B'}\pm\mu^B\bar\mu^{B'})$ is $\delta
_e$-divergence free on ${\cal S}$ for purely imaginary/real 
eigenvalue $\alpha$. The vector fields $k^a$ and $z^a_\pm$ are 
vanishing for eigenspinors with definite chirality, because then 
the null vectors $\lambda^A\bar\mu^{A'}$, $\lambda^A\bar\lambda
^{A'}$ and $\mu^A\bar\mu^{A'}$ are all orthogonal to ${\cal S}$. 

Similarly, the formal substitution $Q^A{}_{EE'B}=0$ in (\ref{eq:2.15}) 
yields that $\xi^e=\phi^A\delta^e\lambda_A+\lambda^A\delta^e\phi
_A$ is a $\delta_e$-divergence-free complex tangent vector field on 
${\cal S}$ if $-2\delta_A{}^{A'}\delta_{A'}{}^B\lambda_B=\alpha^2
\lambda_A$ and $-2\delta_A{}^{A'}\delta_{A'}{}^B\phi_B=\alpha^2
\phi_A$ hold for some function $\alpha:{\cal S}\rightarrow{\mathbb 
C}$, e.g. when $\lambda_A$ and $\phi_A$ are Weyl eigenspinors with 
the {\em same} eigenvalue. If $\lambda_A$ and $\phi_A$ have the 
same, definite chirality, then the vector field $\xi^a$ itself is 
vanishing. 

The analysis of subsection \ref{sub-2.4} can be repeated to obtain 
criteria for the reality of the eigenvalue $\alpha$. The only 
difference is that the Hermitian scalar product on $C^\infty({\cal 
S},{\mathbb S}^A)$ should be defined by 

\begin{equation}
h\bigl[\lambda_A,\bar\omega_{A'}\bigr]:=\frac{1}{4\pi G}\oint_{\cal 
S}\bar\gamma^{A'B'}\bar\omega_{A'}\delta_{B'}{}^B\lambda_B {\rm d}
{\cal S}. \label{eq:3.5}
\end{equation}
Then, for $h[\lambda_A,\bar\lambda_{A'}]\not=0$, the non-zero 
$\alpha$ is real iff $h[\lambda_A,\bar\lambda_{A'}]=-h[\mu_A,\bar
\mu_{A'}]$; and it is purely imaginary iff $h[\lambda_A,\bar\lambda
_{A'}]=h[\mu_A,\bar\mu_{A'}]$. However, for eigenspinors with 
definite chirality the norm $h[\lambda_A,\bar\lambda_{A'}]$ is 
always zero, and hence the corresponding eigenvalue may in principle 
be a (not necessarily real or purely imaginary) complex number.

%%%%%%%%%%%%%%%%%%%%%%%%%%%%%%%%%%%%%%%%%%%%%%%%%%%%%%%%%%%%%%%%%%%%%

\section{On round spheres}
\label{sec-4}

\subsection{The spectrum}
\label{sub-4.1}

Let ${\cal S}$ be a round sphere of radius $r$; i.e. ${\cal S}$ is 
a transitivity surface of the rotation group in a spherically 
symmetric spacetime, whose radius is defined by $4\pi r^2:={\rm 
Area}({\cal S})$. Then the connection in the normal bundle is flat, 
and by an appropriate boost gauge choice $\rho,\rho'={\rm const}$ 
can be achieved. To solve (\ref{eq:2.9}), it could be a good 
strategy to solve (\ref{eq:3.3}) first, and return to (\ref{eq:2.9}) 
later. We expand the spinor components $\lambda_0$ and $\lambda_1$ 
in terms of the spin weighted spherical harmonics: 

\begin{equation}
\lambda_0=\sum_{j=\frac{1}{2}}^\infty\sum_{m=-j}^jc_0^{j,m}\,{}
 _{\frac{1}{2}}Y_{jm}, \hskip 20pt
\lambda_1=\sum_{j=\frac{1}{2}}^\infty\sum_{m=-j}^jc_1^{j,m}\,{}
 _{-\frac{1}{2}}Y_{jm}, \label{eq:4.1}
\end{equation}
where $j=\frac{1}{2},\frac{3}{2},\frac{5}{2},...$ and $m=-j,-j+1,
...,j$; and $c^{j,m}_0$ and $c^{j,m}_1$ are complex constants. Since 
(see. e.g. \cite{PRI}) 

\begin{eqnarray}
{\edth}\,{}_sY_{jm}\!\!\!\!&=\!\!\!\!&-\frac{1}{\sqrt{2}r}\sqrt{
 \bigl(j+s+1\bigr)\bigl(j-s\bigr)}\,{}_{s+1}Y_{jm}, 
 \label{eq:4.2.a} \\
{\edth}'\,{}_sY_{jm}\!\!\!\!&=\!\!\!\!&\frac{1}{\sqrt{2}r}\sqrt{
 \bigl(j-s+1\bigr)\bigl(j+s\bigr)}\,{}_{s-1}Y_{jm}, \label{eq:4.2.b}
\end{eqnarray}
the eigenvalue equations (\ref{eq:3.3}) reduce to 

\begin{equation}
\sum_{j,m}c_0^{j,m}\Bigl(\alpha^2-\frac{1}{r^2}\bigl(j+\frac{1}{2}
 \bigr)^2\Bigr)\,{}_{\frac{1}{2}}Y_{jm}=0, \hskip 10pt 
\sum_{j,m}c_1^{j,m}\Bigl(\alpha^2-\frac{1}{r^2}\bigl(j+\frac{1}{2}
 \bigr)^2\Bigr)\,{}_{-\frac{1}{2}}Y_{jm}=0. \label{eq:4.3}
\end{equation}
Because of the completeness of the spherical harmonics these imply 
that the eigenvalues $\alpha^2$ are {\em discrete}: 

\begin{equation}
\alpha^2=\frac{n^2}{r^2}, \hskip 20pt n:=j+\frac{1}{2}\in{\mathbb N},  
\label{eq:4.4}
\end{equation}
and that, for a given allowed $\alpha$ and the corresponding $j$, 

\begin{equation}
\lambda_0=\sum_{m=-j}^jc_0^m\,{}_{\frac{1}{2}}Y_{jm}, \hskip 20pt
\lambda_1=\sum_{m=-j}^jc_1^m\,{}_{-\frac{1}{2}}Y_{jm} \label{eq:4.5}
\end{equation}
for some complex constants $c_0^m$ and $c_1^m$. Thus the space of 
the right/left handed Weyl eigenspinors is spanned by $c_0^m$ and 
$c_1^m$, respectively, and hence the dimension of these spaces is 
$2j+1=2n$. Then the eigenvalues can also be expressed by the 
(constant) scalar curvature $R$ of the intrinsic geometry of ${\cal 
S}$ as $\alpha^2=n^2\frac{1}{2}R$. Therefore, the first eigenvalue 
already contains all the information about (i.e. the only observable 
of) the geometry of ${\cal S}$. 
The eigenvalues of the $\delta_{AA'}$-Dirac operator are $\alpha=
\pm n/r$, and the components of the corresponding primed Weyl spinor 
are $\bar\mu_{0'}=\mp{\rm i}\sum_mc^m_0{}_{-\frac{1}{2}}Y_{jm}$ and 
$\bar\mu_{1'}=\mp{\rm i}\sum_mc^m_1{}_{\frac{1}{2}}Y_{jm}$. 
We note that for round spheres $\delta_e$-constant Hermitian metrics 
$G_{AA'}$ do exist, and the present particular results can be 
compared with the general ones obtained in pure Riemannian geometry: 
the spectrum (\ref{eq:4.4}) saturates the inequalities of 
\cite{TFr,Hi86,Hi95,Ba92,TF00,FK}. 

By (\ref{eq:2.12.a})-(\ref{eq:2.12.b}) the eigenvalue problem for the 
operators $-2\Delta_A{}^{A'}\Delta_{A'}{}^B$ and $\Delta_{A'}{}^A$ 
can be solved in an analogous way. The structure of the unprimed Weyl 
spinor part of the eigenspinors is similar to that given by 
(\ref{eq:4.5}). The only difference is that the primed Weyl spinor 
part change slightly and the eigenvalues are shifted by a term built 
from the convergences: 

\begin{equation}
\alpha^2=\frac{1}{r^2}\bigl(n^2+2r^2\rho\rho'\bigr), \hskip 25pt 
n:=j+\frac{1}{2}\in{\mathbb N}. \label{eq:4.6}
\end{equation}
In contrast to (\ref{eq:4.4}) this can be zero or even negative in 
certain special geometries. For example, in spacetimes with constant 
curvature the two convergences can be chosen to be $\rho=-\frac{1}{r}
\sqrt{1-\frac{1}{3}\Lambda r^2}$ and $\rho'=\frac{1}{2r}\sqrt{1-
\frac{1}{3}\Lambda r^2}$, $\Lambda\in{\mathbb R}$, and hence $1+2r^2
\rho\rho'=\frac{1}{3}\Lambda r^2$. In Minkowski spacetime ($\Lambda
=0$) $\alpha^2$ is zero, while in the anti-de Sitter spacetime 
($\Lambda<0$) it is negative (i.e. $\alpha$ is purely {\em imaginary}) 
for $n=1$. (For a more general discussion of the kernel of the 
$\Delta_{AA'}$-Dirac operator on round spheres, see \cite{Sz01}.) 
Remarkably enough, by (\ref{eq:4.6}) the first eigenvalue is connected 
with the Hawking quasi-local mass: $\alpha_1^2=2GE_H({\cal S})/r^3$. 
Therefore, the first two eigenvalues of $-2\Delta_A{}^{A'}\Delta_{A'}
{}^B$, or the first eigenvalue of $-2\delta_A{}^{A'}\delta_{A'}{}^B$ 
and of $-2\Delta_A{}^{A'}\Delta_{A'}{}^B$ give the only two 
non-trivial gauge invariant observables of the 2-surface in spacetime, 
namely the scalar curvature $R$ and the length $8\rho\rho'$ of the 
mean curvature vector, or, in other combination, the area of the 
surface and the Hawking energy. 

The components of the primed Weyl spinor part $\bar\mu_{A'}$ of the 
eigenspinors of the $\Delta_e$-Dirac operator can be calculated 
easily from (\ref{eq:2.8}). Since, however, in the present paper 
primarily we are interested in the $\delta_e$-divergence free vector 
fields, we do not need them in what follows.

\subsection{The divergence free vector fields}
\label{sub-4.2}

For given $j$ the $\delta_e$-divergence free vector field $k^a=\Pi
^a_b\lambda^B\bar\mu^{B'}$ takes the form 

\begin{equation*}
k^a=\pm{\rm i}\Bigl(\sum_{k,m}c^k_0c^m_1{}_{-\frac{1}{2}}
Y_{jk}{}_{-\frac{1}{2}}Y_{jm}\Bigr)m^a\pm{\rm i}\Bigl(\sum_{k,m}
c^k_0c^m_1{}_{\frac{1}{2}}Y_{jk}{}_{\frac{1}{2}}Y_{jm}\Bigr)\bar m^a, 
\end{equation*}
where $m^a=\frac{1}{\sqrt{2}r}(1+\zeta\bar\zeta)(\frac{\partial}{
\partial\bar\zeta})^a$. 
In particular, for $j=\frac{1}{2}$ the explicit form of this vector 
field in the standard complex stereographic coordinates $(\zeta,\bar
\zeta)$ is 

\begin{eqnarray}
\pm k^a=\!\!\!\!&-\!\!\!\!&\frac{\rm i}{2\pi}\frac{1}{1+\zeta\bar
 \zeta}\Bigl(c^{\frac{1}{2}}_0c^{\frac{1}{2}}_1-\bigl(c^{\frac{1}{2}}
 _0c^{-\frac{1}{2}}_1+c^{-\frac{1}{2}}_0c^{\frac{1}{2}}_1\bigr)\bar
 \zeta+c^{-\frac{1}{2}}_0c^{-\frac{1}{2}}_1\bar\zeta^2\Bigr)m^a-
 \nonumber \\
\!\!\!\!&-\!\!\!\!&\frac{\rm i}{2\pi}\frac{1}{1+\zeta\bar\zeta}\Bigl(
 c^{\frac{1}{2}}_0c^{\frac{1}{2}}_1\zeta^2+\bigl(c^{\frac{1}{2}}_0c
 ^{-\frac{1}{2}}_1+c^{-\frac{1}{2}}_0c^{\frac{1}{2}}_1\bigr)\zeta+c
 ^{-\frac{1}{2}}_0c^{-\frac{1}{2}}_1\Bigr)\bar m^a. \label{eq:4.8}
\end{eqnarray}
Since there are no harmonic forms on spheres, by the Hodge 
decomposition theorem the divergence-free $rk^a$ can always be 
written as $\varepsilon^{ab}\delta_bF$ for some function $F:{\cal S}
\rightarrow{\mathbb C}$. A short calculation yields that this is 
indeed the case with 

\begin{equation}
F=\mp\frac{r^2}{\sqrt{2}\pi}\frac{c^{-\frac{1}{2}}_0c^{-\frac{1}{2}}
_1\bar\zeta-c^{\frac{1}{2}}_0c^{\frac{1}{2}}_1\zeta+\bigl(c^{\frac{1}
{2}}_0c^{-\frac{1}{2}}_1+c^{-\frac{1}{2}}_0c^{\frac{1}{2}}_1\bigr)
\zeta\bar\zeta}{1+\zeta\bar\zeta}, \label{eq:4.9}
\end{equation}
where the irrelevant constant of integration has been chosen to be 
zero. $k^a$ (and hence the function $F$ also) is real precisely when 

\begin{equation}
c^{-\frac{1}{2}}_0c^{-\frac{1}{2}}_1=a+{\rm i}b, \hskip 15pt
c^{\frac{1}{2}}_0c^{\frac{1}{2}}_1=-a+{\rm i}b, \hskip 15pt
c^{\frac{1}{2}}_0c^{-\frac{1}{2}}_1+c^{-\frac{1}{2}}_0c
^{\frac{1}{2}}_1=2c \label{eq:4.9a}
\end{equation}
for some real constants $a$, $b$ and $c$. If we write $c^{\frac{1}
{2}}_0c^{-\frac{1}{2}}_1=:c+d+{\rm i}e$ for some real $d$ and $e$, 
and hence $c^{-\frac{1}{2}}_0c^{\frac{1}{2}}_1=c-d-{\rm i}e$, then 
$-(a^2+b^2)=(c^{\frac{1}{2}}_0c^{\frac{1}{2}}_1)(c^{-\frac{1}{2}}
_0c^{-\frac{1}{2}}_1)=(c^{\frac{1}{2}}_0c^{-\frac{1}{2}}_1)(c^{-
\frac{1}{2}}_0c^{\frac{1}{2}}_1)=c^2-d^2+e^2-{\rm i}2ed$. Since 
its left hand side is nonpositive, the right hand side must also 
be real and nonpositive, implying that $e=0$, and hence that $d^2
=a^2+b^2+c^2$. 

Since ${\cal S}$ is two dimensional, in the case of real $F$ the 
level sets $F={\rm const}$ are precisely the integral curves of 
the vector field $k^a$, and $F$ is given explicitly by 

\begin{equation}
F=\mp\frac{r^2}{\sqrt{2}\pi}\frac{a\bigl(\zeta+\bar\zeta\bigr)+
 {\rm i}b\bigl(\bar\zeta-\zeta\bigr)+2c\zeta\bar\zeta}{1+\zeta\bar
 \zeta}={\rm const}\mp\frac{r}{\sqrt{2}\pi}\Bigl(ax+by+cz\Bigr). 
 \label{eq:4.10}
\end{equation}
(Here $x$, $y$ and $z$ are the standard Cartesian coordinates 
defined by $x+{\rm i}y:=2r\zeta/(1+\zeta\bar\zeta)$ and $z:=r(\zeta
\bar\zeta-1)/((1+\zeta\bar\zeta)$.) However, the corresponding level 
sets are just the integral curves of the Killing vectors of the 
metric 2-sphere. In fact, in these coordinates the standard rotation 
Killing 1-forms are 

\begin{eqnarray}
K^{23}_a\!\!\!\!&=\!\!\!\!&\frac{\rm i}{\sqrt2}\frac{r}{1+\zeta\bar
 \zeta}\Bigl(\bigl(1-\bar\zeta^2\bigr)m_a-\bigl(1-\zeta^2\bigr)\bar 
 m_a\Bigr), \nonumber \\
K^{31}_a\!\!\!\!&=\!\!\!\!&\frac{1}{\sqrt2}\frac{r}{1+\zeta\bar\zeta}
 \Bigl(\bigl(1+\bar\zeta^2\bigr)m_a+\bigl(1+\zeta^2\bigr)\bar m_a
 \Bigr), \label{eq:K}\\
K^{12}_a\!\!\!\!&=\!\!\!\!&\frac{{\rm i}r\sqrt{2}}{1+\zeta\bar\zeta}
 \Bigl(\bar\zeta m_a-\zeta\bar m_a\Bigr). \nonumber
\end{eqnarray}
Comparing these with (\ref{eq:4.8}), we find that 

\begin{eqnarray}
rk_a=\pm\frac{1}{2\sqrt{2}\pi}\Bigl(\bigl(c^{-\frac{1}{2}}_0c
 ^{-\frac{1}{2}}_1-c^{\frac{1}{2}}_0c^{\frac{1}{2}}_1\bigr)K^{23}_a
 \!\!\!\!&-\!\!\!\!&{\rm i}\bigl(c^{\frac{1}{2}}_0c^{\frac{1}{2}}_1+
 c^{-\frac{1}{2}}_0c^{-\frac{1}{2}}_1\bigr)K^{31}_a+\nonumber \\
\!\!\!\!&+\!\!\!\!&\bigl(c^{\frac{1}{2}}_0c^{-\frac{1}{2}}_1+c^{-
 \frac{1}{2}}_0c^{\frac{1}{2}}_1\bigr)K^{12}_a\Bigr); \label{eq:4.11}
\end{eqnarray}
i.e. {\em on round spheres of radius $r$ the $\delta_e$-divergence 
free vector field $k^a$ built from the $j=\frac{1}{2}$ eigenspinors 
is $1/r$-times a complex combination of the rotation Killing vectors.} 

By a similar analysis one can determine the vector fields $k^a$ for 
all $j>\frac{1}{2}$. In particular, for $j=\frac{3}{2}$ there are 
precisely {\em ten} such complex independent vector fields. The 
functions $F$ corresponding to the independent real vector fields 
$rk^a$ are cubic expressions of the Cartesian coordinates $x$, $y$ 
and $z$, divided by $r$, e.g. $(x^3-3xy^2)/r$, $xyz/r$, ... etc. 
However, their significance and geometric meaning are still not quite 
clear. 

Since the eigenvalues $\alpha^2$ are real, the vector fields $z
_-^a$ are $\delta_e$ divergence free. Repeating the analysis above, 
a direct calculation shows that for $j=\frac{1}{2}$ these vector 
fields have the form $\frac{1}{r}(AK^{23}_a+BK^{31}_a+CK^{12}_a)$ 
for some {\em real} constants $A$, $B$ and $C$; i.e. {\em for $j=
\frac{1}{2}$ the vector field $rz^a_-$ is a real linear combination 
of the three rotation Killing vectors}. 
For $j=\frac{1}{2}$ the vector field $\xi^a$ is also proportional 
to the rotation Killing vector fields. Since, however, $\xi^a$ is 
built from the {\em first derivative} of the Weyl spinor parts of 
the eigenspinor, it scales with a different power of the radius: 
$\xi_a=\frac{1}{r^2}(AK^{23}_a+BK^{31}_a+CK^{12}_a)$ with appropriate 
(complex) constants $A$, $B$ and $C$. We continue the discussion of 
$k^a$ and $z^a_-$ in a more general context in subsections 
\ref{sub-5.3.1} and \ref{sub-5.3.2}, respectively.

%%%%%%%%%%%%%%%%%%%%%%%%%%%%%%%%%%%%%%%%%%%%%%%%%%%%%%%%%%%%%%%%%%

\section{On large spheres near the null infinity}
\label{sec-5}

\subsection{Asymptotically flat spacetimes}
\label{sub-5.1}

Let the spacetime be asymptotically flat at future null infinity, and 
let $(u,r,\zeta,\bar\zeta)$ be a Bondi-type coordinate system (see 
e.g. \cite{NT}). Then the standard edth operators, acting on $(p,q)$ 
type scalar $f$, take the form ${\edth}f=P(\partial f/\partial\bar
\zeta)+Q(\partial f/\partial\zeta)-p\beta f+q\bar\beta'f$ and 
${\edth}'f=P(\partial f/\partial\zeta)+\bar Q(\partial f/\partial\bar
\zeta)+p\beta'f-q\bar\beta f$, respectively. (For the definition of 
the GHP spin coefficients see \cite{GHP,PRI}.) The GHP spin frame is 
chosen such that $o^A$ is parallelly propagated along the null geodesic 
generators of the null hypersurfaces $u={\rm const}$, and $\iota^A$ 
is chosen such that $m^a:=o^A\bar\iota^{A'}$ and $\bar m^a:=\iota^A
\bar o^{A'}$ be tangents to the spacelike 2-surfaces $u={\rm const}$, 
$r={\rm const}$. In this case $\bar\beta'=\beta-\tau$. In these 
coordinates and spin frame in an Einstein--Maxwell spacetime the 
asymptotic form of the functions $P$ and $Q$ and some of the spin 
coefficients that we need are well known \cite{Sh} to be given by 
$P=\frac{1}{r\sqrt{2}}(1+\zeta\bar\zeta)+O(r^{-3})$, 
$Q=-\frac{1}{r^2\sqrt{2}}(1+\zeta\bar\zeta)\sigma^0+O(r^{-4})$, 
$\beta=-\frac{1}{r2\sqrt{2}}\zeta-\frac{1}{r^22\sqrt{2}}\bar\zeta
\sigma^0+O(r^{-3})$ and 
$\tau=\frac{1}{r^2}{}_0{\edth}'\sigma^0-\frac{1}{r^3}(2\sigma^0
{}_0{\edth}\bar\sigma^0+\psi^0_1)+O(r^{-4})$. 
Here ${}_0{\edth}$ and ${}_0{\edth}'$ denote the edth operators on 
the metric, unit sphere, $\sigma^0$ is the asymptotic shear (i.e. 
$\sigma=\frac{1}{r^2}\sigma^0+O(r^{-4})$), the dot denotes 
differentiation with respect to $u$, and $\psi^0_1$, $\psi^0_2$ are 
the leading terms in the asymptotic expansion of the Weyl spinor 
components $\Psi_1$ and $\Psi_2$, respectively (see e.g. 
\cite{NT,PRII}). 
In addition, the Weyl spinor components satisfy 

\begin{eqnarray}
\dot\psi^0_1\!\!\!\!&=\!\!\!\!&{}_0{\edth}\psi^0_2-2\sigma^0{}_0
 {\edth}\dot{\bar\sigma}{}^0+4G\varphi^0_1\bar\varphi^0_2, 
 \label{eq:5.0.1a}\\
\dot\psi^0_2\!\!\!\!&=\!\!\!\!&-{}_0{\edth}^2\dot{\bar\sigma}{}^0_2
 -\sigma^0\ddot{\bar\sigma}{}^0+2G\varphi^0_2\bar\varphi^0_2, 
 \label{eq:5.0.1b}\\
\psi^0_2-\bar\psi^0_2\!\!\!\!&=\!\!\!\!&{}_0{\edth}'{}^2\sigma
 ^0-{}_0{\edth}{}^2\bar\sigma^0+\bar\sigma^0\dot\sigma^0-\dot{\bar
 \sigma}{}^0\sigma^0. \label{eq:5.0.1c}
\end{eqnarray}
Here $\varphi^0_n$ are the leading terms in the asymptotic expansion 
of the Maxwell spinor components: $\varphi_n=r^{n-3}\varphi^0_n+O(
r^{n-4})$. We will use these formulae in subsection \ref{sub-5.4.3}.

\subsection{The eigenvalue problem on large spheres}
\label{sub-5.2}

In the Bondi type coordinate system the spacelike 2-surfaces $u=
{\rm const}$, $r={\rm const}$ for large enough $r$ are called large 
spheres (and will be denoted by ${\cal S}_r$), and our aim is to 
solve the eigenvalue problem (\ref{eq:3.3}) on these surfaces 
asymptotically. To do so, let us write the components $\lambda
_{\uA}$ of the spinor field $\lambda_A$ in the GHP spin frame 
$\varepsilon^A_{\uA}:=\{o^A,\iota^A\}$, ${\uA}=0,1$, as $\lambda
_{\uA}=:\lambda^{(0)}_{\uA}+\frac{1}{r}\lambda^{(1)}_{\uA}+...$ 
and, similarly, we expand the eigenvalue as $r^2\alpha^2=:\alpha
^2_{(0)}+\frac{1}{r}\alpha^2_{(1)}+...$. (Thus we adopt the 
convention that an index between parentheses, e.g. 0 or 1 here, 
is referring to the order of approximation.) Then substituting 
all these into equation (\ref{eq:3.3}) we obtain 

\begin{eqnarray}
2\,{}_0{\edth}\,{}_0{\edth}'\lambda^{(0)}_0+\alpha^2_{(0)}\lambda
 ^{(0)}_0\!\!\!\!&=\!\!\!\!&0 \label{eq:5.1.a} \\
2\,{}_0{\edth}\,{}_0{\edth}'\lambda^{(1)}_0+\alpha^2_{(0)}\lambda
 ^{(1)}_0\!\!\!\!&=\!\!\!\!&-\alpha^2_{(1)}\lambda^{(0)}_0+2{}_0
 {\edth}'\bigl(\sigma^0\,{}_0{\edth}'\lambda^{(0)}_0\bigr)+2{}_0
 {\edth}^2\bigl(\bar\sigma^0\lambda^{(0)}_0\bigr) \label{eq:5.1.b}
\end{eqnarray}
and 

\begin{eqnarray}
2\,{}_0{\edth}'\,{}_0{\edth}\lambda^{(0)}_1+\alpha^2_{(0)}\lambda^{(0)}
 _1\!\!\!\!&=\!\!\!\!&0 \label{eq:5.2.a} \\
2\,{}_0{\edth}'\,{}_0{\edth}\lambda^{(1)}_1+\alpha^2_{(0)}\lambda^{(1)}
 _1\!\!\!\!&=\!\!\!\!&-\alpha^2_{(1)}\lambda^{(0)}_1+2{}_0{\edth}'
 \bigl(\sigma^0\,{}_0{\edth}'\lambda^{(0)}_1\bigr)+2\bar\sigma^0\,{}_0
 {\edth}^2\lambda^{(0)}_1. \label{eq:5.2.b}
\end{eqnarray}
The zeroth order equations (\ref{eq:5.1.a}) and (\ref{eq:5.2.a}) are 
just the eigenvalue equations (\ref{eq:3.3}) on the unit sphere. Thus 
in the zeroth order the eigenvalues are $\alpha^2_{(0)}=n^2$, $n:=j+
\frac{1}{2}\in{\mathbb N}$, and the components $\lambda_0^{(0)}$ and 
$\lambda_1^{(0)}$ of the eigenspinors are given explicitly by 
(\ref{eq:4.5}), where in general the expansion coefficients are still 
arbitrary functions of the retarded time coordinate $u$. Since the 
structure of the left hand side of equation (\ref{eq:5.1.b}) is similar 
to the homogeneous (\ref{eq:5.1.a}), and that of (\ref{eq:5.2.b}) to 
the homogeneous (\ref{eq:5.2.a}), the general solution of the 
homogeneous equations can be added to the first order corrections 
$\lambda^{(1)}_0$ and $\lambda^{(1)}_1$, yielding an ambiguity in the 
order $O(r^{-1})$. 

Next let us calculate the first-order correction to the zeroth-order 
expression of the {\em first eigenvalue}, $\alpha^2_{(0)}=1$, and the 
corresponding eigenspinors. Since 
$\lambda^{(0)}_0={}_{(0)}c_0^{\frac{1}{2}}\,{}_{\frac{1}{2}}
Y_{\frac{1}{2}\frac{1}{2}}+{}_{(0)}c_0^{-\frac{1}{2}}\,
{}_{\frac{1}{2}}Y_{\frac{1}{2}-\frac{1}{2}}$ and 
$\lambda^{(0)}_1={}_{(0)}c_1^{\frac{1}{2}}\,{}_{-\frac{1}{2}}
Y_{\frac{1}{2}\frac{1}{2}}+{}_{(0)}c_1^{-\frac{1}{2}}\,{}_{-
\frac{1}{2}}Y_{\frac{1}{2}-\frac{1}{2}}$, where the coefficients 
${}_{(0)}c^m_{\uA}$ are functions of $u$, one has 
${}_0{\edth}\lambda^{(0)}_0=0$, 
${}_0{\edth}'\lambda^{(0)}_1=0$, 
${}_0{\edth}'{}^2\lambda^{(0)}_0=0$, 
${}_0{\edth}^2\lambda^{(0)}_1=0$, and 
${}_0{\edth}{}_0{\edth}'\lambda_0^{(0)}=-\frac{1}{2}\lambda^{(0)}_0$, 
${}_0{\edth}'{}_0{\edth}\lambda_1^{(0)}=-\frac{1}{2}\lambda^{(0)}_1$. 
Using these the equations (\ref{eq:5.1.b}) and (\ref{eq:5.2.b}), 
respectively, reduce to 

\begin{eqnarray}
2\,{}_0{\edth}\,{}_0{\edth}'\lambda^{(1)}_0+\lambda^{(1)}_0\!\!\!\!&=
 \!\!\!\!&-\alpha^2_{(1)}\lambda^{(0)}_0+2\bigl({}_0{\edth}'\sigma^0
 \bigr)\bigl(\,{}_0{\edth}'\lambda^{(0)}_0\bigr)+2\bigl({}_0{\edth}^2
 \bar\sigma^0\bigr)\lambda^{(0)}_0, \label{eq:5.4} \\
2\,{}_0{\edth}'\,{}_0{\edth}\lambda^{(1)}_1+\lambda^{(1)}_1\!\!\!\!&=
 \!\!\!\!&-\alpha^2_{(1)}\lambda^{(0)}_1. \label{eq:5.5}
\end{eqnarray}
To solve (\ref{eq:5.5}) let us write $\lambda^{(1)}_1=:\sum_{j,m}\,
{}_{(1)}c^{jm}_1{}_{-\frac{1}{2}}Y_{jm}$ with arbitrary functions 
${}_{(1)}c^{jm}_1={}_{(1)}c^{jm}_1(u)$. We obtain that $\alpha_{(1)}
=0$ (i.e. in particular the first eigenvalue is real in the first 
two orders), and all the 
expansion coefficients ${}_{(1)}c_1^{jm}$ must be zero for $j\geq
\frac{3}{2}$. Hence $\lambda^{(1)}_1$ is a combination only of the 
two $-\frac{1}{2}$ spin weighted spherical harmonics ${}_{-\frac{1}
{2}}Y_{\frac{1}{2}\pm\frac{1}{2}}$, i.e. the correction to $\lambda
^{(0)}_1$ has the structure similar to that of $\lambda^{(0)}_1$ 
itself. Thus $\lambda^{(1)}_1$, as the first-order correction to 
the zeroth-order eigenspinor, represents only a pure (gauge) 
ambiguity. 

To solve (\ref{eq:5.4}) let us observe that the operator ${\edth}$ 
acting on scalars with positive spin weight on topological 2-spheres 
is surjective (see e.g. \cite{PRI,FSz}). Thus the asymptotic shear 
$\sigma^0$ can always be derived from an appropriate {\em complex} 
scalar $S$ of spin weight zero (indeed, of type $(1,1)$): $\sigma^0
={}_0{\edth}^2S$, where the ambiguities in $S$ are the elements of 
$\ker{}_0{\edth}^2$, and, on any given surface $u={\rm const}$, they 
form a four complex dimensional space. Then taking into account 
$\alpha_{(1)}=0$ and the commutator $({}_0{\edth}{}_0{\edth}'-{}_0
{\edth}'{}_0{\edth})f=-\frac{1}{2}(p-q)f$ acting on the $(p,q)$ type 
scalar $f$, (\ref{eq:5.4}) can be written into the form 

\begin{equation}
{}_0{\edth}'\Bigl({}_0{\edth}\lambda^{(1)}_0-{}_0{\edth}^2\bigl(S-
\bar S\bigr)\bigl({}_0{\edth}'\lambda^{(0)}_0\bigr)-\bigl({}_0
{\edth}'{}_0{\edth}^2\bar S\bigr)\lambda^{(0)}_0\Bigr)=0. 
\label{eq:5.6}
\end{equation}
However, $\dim\ker{\edth}'=0$ for positive spin weight scalars on 
topological 2-spheres, and hence the expression between the big 
parentheses itself is zero. Nevertheless, using the commutator of 
${}_0{\edth}$ and ${}_0{\edth}'$ above, this expression can also 
be rewritten as a pure ${}_0{\edth}$-derivative, and hence we have 

\begin{equation}
{}_0{\edth}\Bigl(\lambda^{(1)}_0-{}_0{\edth}\bigl(S-\bar S\bigr)
\bigl({}_0{\edth}'\lambda^{(0)}_0\bigr)-\frac{1}{2}\bigl(S+\bar S
\bigr)\lambda^{(0)}_0-\bigl({}_0{\edth}{}_0{\edth}'\bar S\bigr)
\lambda^{(0)}_0\Bigr)=0. \label{eq:5.7}
\end{equation}
Finally, since $\dim\ker{\edth}=2$ for ${\edth}$ acting on scalars 
of spin weight $\frac{1}{2}$ on topological 2-spheres, from 
(\ref{eq:5.7}) we can deduce that 

\begin{equation}
\lambda^{(1)}_0={}_0{\edth}\bigl(S-\bar S\bigr)\bigl({}_0{\edth}'
\lambda^{(0)}_0\bigr)+\bigl({}_0{\edth}{}_0{\edth}'\bar S+\frac{1}{2}
\bigl(S+\bar S\bigr)\bigr)\lambda^{(0)}_0+\lambda, \label{eq:5.8}
\end{equation}
where $\lambda$ is an arbitrary spin weight $\frac{1}{2}$ solution 
of ${}_0{\edth}\lambda=0$. Thus, for $j=\frac{1}{2}$, equations 
(\ref{eq:5.1.a})--(\ref{eq:5.2.b}) can be solved {\em explicitly} in 
terms of the complex potential $S$ for the asymptotic shear $\sigma
^0$. 

The ambiguities in (\ref{eq:5.8}), coming from the ambiguity of the 
potential $S$ (i.e. formally from the non-triviality of the kernel of 
${}_0{\edth}^2$ acting on zero spin weight scalars) and the 
ambiguity of the solution of the inhomogeneous equation 
(\ref{eq:5.1.b}) (i.e. from the non-triviality of the kernel of ${}_0
{\edth}$ acting on $\frac{1}{2}$ spin weight scalars), can be 
summarized as 

\begin{equation}
\lambda^{(1)}_0\mapsto\lambda^{(1)}_0+\sum_{m=-\frac{1}{2}}
^{\frac{1}{2}}{}_{(1)}c_0^m\,{}_{\frac{1}{2}}Y_{\frac{1}{2}m},
\label{eq:5.9}
\end{equation}
where ${}_{(1)}c_0^{\pm\frac{1}{2}}$ are arbitrary complex functions 
of $u$. Here we used the expansion of the products of spin weighted 
spherical harmonics of the form ${}_0Y_{1M}\,{}_{\frac{1}{2}}Y
_{\frac{1}{2}m}$ and ${}_1Y_{1M}\,{}_{-\frac{1}{2}}Y_{\frac{1}{2}m}$ 
in terms of ${}_{\frac{1}{2}}Y_{\frac{1}{2}m}$ and ${}_{\frac{1}{2}}
Y_{\frac{3}{2}m'}$, which can easily be derived by direct calculation 
from the explicit expression of the harmonics given e.g. in \cite{St}. 
Though in these expansions the spherical harmonics ${}_{\frac{1}{2}}
Y_{\frac{3}{2}m}$ do appear, all these are canceled from (\ref{eq:5.8}). 
Thus the ambiguities in $S$ described by the ${}_0Y_{1m}$ spherical 
harmonics are all canceled from the eigenspinors. The remaining scalar 
ambiguity in $S$ yields an ambiguity in $\lambda^{(1)}_0$ similar to 
that coming from the solutions of the homogeneous equation. Hence all 
these can be parameterized only by two (rather than the originally 
expected six) complex functions of $u$. 

Although there is no {\em canonical} isomorphism between the spaces of 
the eigenspinors on two different surfaces even if these spaces can 
be mapped to each other isomorphically, in the asymptotically flat 
context we can introduce a natural {\em equivalence} between the space 
of the eigenspinors on the large spheres, say ${\cal S}'$ and ${\cal 
S}''$, at different retarded times $u=u'$ and $u=u''$. Namely, {\em 
we require that the zeroth-order solutions be independent of the 
retarded time coordinate $u$}, i.e. we choose the expansion 
coefficients ${}_{(0)}c^m_{\uA}$ constant. 
(It is easy to see that this requirement can be extended to the case 
when the cuts of $\mathscr{I}^+$ that the large spheres define, say 
$\hat{\cal S}'$ and $\hat{\cal S}''$, are related to each other only 
by a proper BMS {\em supertranslation}. The basis of this equivalence 
is the fact that (1) the spinor bases are uniquely determined on the 
large spheres ${\cal S}'$ and ${\cal S}''$ such that these bases are 
related via the spinor constituent of the null geodesic generators 
of $\mathscr{I}^+$; (2) the coordinates $(\zeta',\bar\zeta')$ on 
${\cal S}'$ determine $(\zeta'',\bar\zeta'')$ on ${\cal S}''$ 
uniquely via the null geodesic generators of $\mathscr{I}^+$; (3) in 
these bases and coordinates the zeroth-order solutions $(\lambda'{}
^{(0)}_0,\lambda'{}^{(0)}_1)$ on ${\cal S}'$ and $(\lambda''{}^{(0)}
_0,\lambda''{}^{(0)}_1)$ on ${\cal S}''$ have the same structure if 
we expand them in terms of the spin weighted spherical harmonics. 
Then we identify $(\lambda'{}^{(0)}_0,\lambda'{}^{(0)}_1)$ with 
$(\lambda''{}^{(0)}_0,\lambda''{}^{(0)}_1)$ precisely when their 
spinor components are combined from the spin weighted spherical 
harmonics by the same complex coefficients. This condition is 
analogous to a choice of a conformal gauge on $\mathscr{I}^+$. Then 
an eigenspinor $\lambda'_A$ on ${\cal S}'$ will be called equivalent 
to the eigenspinor $\lambda''_A$ on ${\cal S}''$ if their zeroth-order 
parts are the same in the sense above.)  

Though this condition does not rule out the $u$-dependence of the 
first order correction terms $\lambda^{(1)}_{\uA}$, this makes the 
whole solution considerably more ``rigid'', and essentially their 
$u$-dependence is already controlled: all of their ambiguities with 
arbitrary $u$-dependence have the form 

\begin{equation}
\sum_{m=-\frac{1}{2}}^{\frac{1}{2}}{}_{(1)}c_0^m(u)\,{}_{\frac{1}{2}}
Y_{\frac{1}{2}m}, \hskip 25pt 
\sum_{m=-\frac{1}{2}}^{\frac{1}{2}}{}_{(1)}c_1^m(u)\,{}_{-\frac{1}{2}}
Y_{\frac{1}{2}m},\label{eq:5.10}
\end{equation}
i.e. they belong to the kernel of ${}_0{\edth}$ and ${}_0{\edth}'$, 
respectively, while the $u$-dependence of the remaining $u$-dependent 
parts (of $\lambda^{(1)}_0$, see (\ref{eq:5.8})) comes only from the 
$u$-dependence of the asymptotic shear. Thus the role of the 
equivalence is to provide a common, {\em universal} (i.e. 
cut-independent) parameter space, namely the space of the zeroth 
order solutions, by means of which the solutions (up to the 
ambiguities (\ref{eq:5.10})) can be parameterized. As we will see in 
subsections \ref{sub-5.4.1} and \ref{sub-5.4.2}, the general 
expression of our 2-surface observable is not sensitive to this 
ambiguity, and this makes it possible to be able to compare angular 
momenta defined at different retarded times in subsection 
\ref{sub-5.4.3}.

\subsection{The $\delta_e$--divergence free vector fields for $j=
\frac{1}{2}$}
\label{sub-5.3}

From $\lambda^{(0)}_{\uA}$ and $\lambda^{(1)}_{\uA}$ we can compute 
the components of $\bar\mu_{A'}$ using (\ref{eq:3.1}) and the 
asymptotic form of the edth operators. For the eigenvalue $\alpha
_{(0)}=\pm1$ we have 

\begin{eqnarray}
\bar\mu_{0'}\!\!\!\!&=\!\!\!\!&\mp{\rm i}\sqrt{2}\Bigl({}_0{\edth}'
 \lambda^{(0)}_0+\frac{1}{r}{}_0{\edth}'\lambda^{(1)}_0-\frac{1}{r}
 \bigl({}_0{\edth}\bar\sigma^0\bigr)\lambda^{(0)}_0\Bigr), 
 \label{eq:5.3.1a} \\
\bar\mu_{1'}\!\!\!\!&=\!\!\!\!&\pm{\rm i}\sqrt{2}\Bigl({}_0{\edth}
 \lambda^{(0)}_1+\frac{1}{r}{}_0{\edth}\lambda^{(1)}_1\Bigr), 
 \label{eq:5.3.1b}
\end{eqnarray}
and in the next two subsections we discuss the $\delta
_e$--divergence-free vector fields $k^a$ and $z^a_-$ built from $\lambda
_A$ and $\bar\mu_{A'}$.

\subsubsection{The vector field $k^a$}
\label{sub-5.3.1}

For the vector field $k^a$ we obtain 

\begin{eqnarray}
\pm k^a=\frac{\rm i}{r}\sqrt{2}\!\!\!\!&\Bigl(\!\!\!\!&{}_0{\edth}'
 \bigl(\lambda^{(0)}_0\lambda^{(0)}_1\bigr)\hat m^a-{}_0{\edth}
 \bigl(\lambda^{(0)}_0\lambda^{(0)}_1\bigr)\bar{\hat m}{}^a\Bigr)+ 
 \nonumber \\
+\frac{\rm i}{r^2}\sqrt{2}\!\!\!\!&\Bigl(\!\!\!\!&\Bigl[\lambda^{(1)}
 _1\bigl({}_0{\edth}'\lambda^{(0)}_0\bigr)+\lambda^{(0)}_1\bigl({}
 _0{\edth}'\lambda^{(1)}_0\bigr)-\lambda^{(0)}_0\lambda^{(0)}_1{}_0
 {\edth}\bar\sigma^0\Bigr]\hat m^a-\nonumber \\
\!\!\!\!&-\!\!\!\!&\Bigl[\lambda^{(1)}_0\bigl({}_0{\edth}\lambda
 ^{(0)}_1\bigr)+\lambda^{(0)}_0\bigl({}_0{\edth}\lambda^{(1)}_1
 \bigr)\Bigr]\bar{\hat m}{}^a\Bigr)+O\bigl(r^{-3}\bigr), 
 \label{eq:5.3.2}
\end{eqnarray}
where $\hat m^a:=\frac{1}{\sqrt{2}}(1+\zeta\bar\zeta)(\partial/
\partial\bar\zeta)^a$, the complex null vector on the unit sphere, 
normalized with respect to the unit sphere metric. Since $rk^a$ is 
divergence free, it can always be written as the dual of the gradient 
of some function $F$: 

\begin{eqnarray}
rk^a\!\!\!\!&=\!\!\!\!&\varepsilon^{ab}\delta_bF=-{\rm i}\Bigl(m^a
 {\edth}'F-\bar m^a{\edth}F\Bigr)= \nonumber \\
\!\!\!\!&=\!\!\!\!&\frac{\rm i}{r^2}\Bigl(\bigl({}_0{\edth}F\bigr)-
 \frac{1}{r}\sigma^0\bigl({}_0{\edth}'F\bigr)\Bigr)\bar{\hat m}{}^a-
 \frac{\rm i}{r^2}\Bigl(\bigl({}_0{\edth}'F\bigr)-\frac{1}{r}\bar
 \sigma^0\bigl({}_0{\edth}F\bigr)\Bigr){\hat m}{}^a. \label{eq:5.3.3}
\end{eqnarray}
Writing the function $F$ as $F=:r^2F^{(-2)}+rF^{(-1)}+O(1)$ and 
comparing (\ref{eq:5.3.2}) with (\ref{eq:5.3.3}) we obtain a system of 
partial differential equations for $F^{(-2)}$ and $F^{(-1)}$. Their 
integrability condition is satisfied by (\ref{eq:5.8}), and the 
solution is 

\begin{eqnarray}
F^{(-2)}=\!\!\!\!&\mp\!\!\!\!&\sqrt{2}\lambda^{(0)}_0\lambda^{(0)}
 _1, \label{eq:5.3.4a} \\
F^{(-1)}=\!\!\!\!&\mp\!\!\!\!&\sqrt{2}\Bigl\{\bigl({}_0{\edth}
 S\bigr){}_0{\edth}'\bigl(\lambda^{(0)}_0\lambda^{(0)}_1\bigr)+
 \bigl({}_0{\edth}'\bar S\bigr){}_0{\edth}\bigl(\lambda^{(0)}_0
 \lambda^{(0)}_1\bigr)- \nonumber \\
\!\!\!\!&-\!\!\!\!&\frac{1}{2}\bigl(S+\bar S\bigr){}_0{\edth}'{}_0
 {\edth}\bigl(\lambda^{(0)}_0\lambda^{(0)}_1\bigr)+\frac{1}{2}
 \bigl(S-\bar S\bigr)\Bigl(\lambda^{(0)}_0\lambda^{(0)}_1+{}_0
 {\edth}'{}_0{\edth}\bigl(\lambda^{(0)}_0\lambda^{(0)}_1\bigr)
 \Bigr)+ \nonumber \\
\!\!\!\!&+\!\!\!\!&\lambda^{(0)}_0\lambda^{(1)}_1+\lambda\lambda
 ^{(0)}_1\Bigr\}, \label{eq:5.3.4b} 
\end{eqnarray}
where the irrelevant constants of integration have been chosen to be 
zero. Apart from the last two (ambiguous) terms in (\ref{eq:5.3.4b}) 
(which have the structure of $F^{(-2)}$) all the terms are proportional 
to $\lambda^{(0)}_0\lambda^{(0)}_1$, which is essentially $F^{(-2)}$. 
Although $r^2F^{(-2)}$ has been given explicitly by (\ref{eq:4.9}), 
here we give another, and, from the points of view of later 
applications, a more useful expression. 

Let us define the real (essentially the ${}_0Y_{1m}$ spherical 
harmonic) functions 

\begin{equation}
t^1=\frac{\zeta+\bar\zeta}{1+\zeta\bar\zeta}, \hskip 20pt
t^2={\rm i}\frac{\bar\zeta-\zeta}{1+\zeta\bar\zeta}, \hskip 20pt
t^3=\frac{\zeta\bar\zeta-1}{1+\zeta\bar\zeta}, \label{eq:5.3.5}
\end{equation}
and introduce the coefficients 

\begin{eqnarray}
R_1\!\!\!\!&:=\!\!\!\!&\frac{1}{4\sqrt{2}\pi}\Bigl({}_{(0)}c^{
 -\frac{1}{2}}_0{}_{(0)}c^{-\frac{1}{2}}_1-{}_{(0)}c^{\frac{1}{2}}_0
 {}_{(0)}c^{\frac{1}{2}}_1\Bigr)=\frac{1}{2\sqrt{2}\pi}a, 
 \label{eq:5.3.6a} \\
R_2\!\!\!\!&:=\!\!\!\!&-\frac{\rm i}{4\sqrt{2}\pi}\Bigl({}
 _{(0)}c^{-\frac{1}{2}}_0{}_{(0)}c^{-\frac{1}{2}}_1+{}_{(0)}c
 ^{\frac{1}{2}}_0{}_{(0)}c^{\frac{1}{2}}_1\Bigr)=\frac{1}{2\sqrt{2}
 \pi}b, \label{eq:5.3.6b} \\
R_3\!\!\!\!&:=\!\!\!\!&\frac{1}{4\sqrt{2}\pi}\Bigl({}_{(0)}c^{
 \frac{1}{2}}_0{}_{(0)}c^{-\frac{1}{2}}_1+{}_{(0)}c^{-\frac{1}{2}}_0
 {}_{(0)}c^{\frac{1}{2}}_1\Bigr)=\frac{1}{2\sqrt{2}\pi}c, 
 \label{eq:5.3.6c}\\
R_0\!\!\!\!&:=\!\!\!\!&\frac{1}{4\sqrt{2}\pi}\Bigl({}_{(0)}c^{\frac{1}
 {2}}_0{}_{(0)}c^{-\frac{1}{2}}_1-{}_{(0)}c^{-\frac{1}{2}}_0{}_{(0)}
 c^{\frac{1}{2}}_1\Bigr)=\frac{1}{2\sqrt{2}\pi}d; \label{eq:5.3.6d}
\end{eqnarray}
which are constant by our requirement imposed at the end of the previous 
subsection. Here the second equalities hold if the reality condition 
(\ref{eq:4.9a}) is imposed, and hence $d$ is determined up to sign by 
the parameters in $R_{\bi}$, ${\bi}=1,2,3$, as $d^2=a^2+b^2+c^2$, i.e. 
formally $(R_0,R_{\bi})$ is a real null vector with respect to the 
constant Lorentzian metric. Then a simple calculation gives 

\begin{equation}
\lambda^{(0)}_0\lambda^{(0)}_1=\sqrt{2}\Bigl(R_0+R_{\bi}t^{\bi}\Bigr), 
\label{eq:5.3.7}
\end{equation}
yielding a parameterization of $F^{(-2)}$ and $F^{(-1)}$ in terms of 
$R_{\bi}$ and $R_0$: 

\begin{eqnarray}
F^{(-2)}=\!\!\!\!&\mp\!\!\!\!&2\Bigl(R_0+R_{\bi}t^{\bi}\Bigr), 
 \label{eq:5.3.8a} \\
F^{(-1)}=\!\!\!\!&\mp\!\!\!\!&2\Bigl(\bigl({}_0{\edth}S\bigr)\bigl(
 {}_0{\edth}'t^{\bi}\bigr)+\bigl({}_0{\edth}'\bar S\bigr)\bigl({}_0
 {\edth}t^{\bi}\bigr)+\frac{1}{2}\bigl(S+\bar S\bigr)t^{\bi}\Bigr)
 R_{\bi}\mp \nonumber \\
\!\!\!\!&\mp\!\!\!\!&\bigl(S-\bar S\bigr)R_0\mp 2\bigl(G_0+G_{\bi}t
 ^{\bi}\bigr),\label{eq:5.3.8b} 
\end{eqnarray}
where the last term of (\ref{eq:5.3.8b}) comes from the last two 
ambiguous terms of (\ref{eq:5.3.4b}) for some (in general complex) 
functions $G_0$ and $G_{\bi}$ of $u$. Introducing the new notation 
$K^{\bi}_a:=\frac{1}{2}\varepsilon^{\bi}{}_{\bj\bk}K^{\bj\bk}_a$ for 
the Killing fields of the metric sphere of radius $r$ and defining 
the antisymmetric matrix $M_{\bi\bj}$ by $R_{\bi}=:\frac{1}{2}
\varepsilon_{\bi}{}^{\bj\bk}M_{\bj\bk}$, the leading-order term in 
$rk_a$ can in fact be written as $K^{\bi\bj}_aM_{\bi\bj}=-2R_{\bi}K
^{\bi}_a$, as it could be expected by (\ref{eq:4.11}). (Here raising 
and lowering of the boldface indices are defined by the {\em 
negative} definite constant metric $\eta_{\bi\bj}:=-\delta_{\bi\bj}$.) 
The vector field $rk^a$ itself is 

\begin{eqnarray}
\mp rk^a=2{\rm i}\!\!\!\!&\Bigl(\!\!\!\!&\bar{\hat m}{}^a\bigl({}_0
 {\edth}t^{\bi}\bigr)-\hat m{}^a\bigl({}_0{\edth}'t^{\bi}\bigr)
 \Bigr)R_{\bi}+ \nonumber \\
+\frac{\rm i}{r}\!\!\!\!&\Bigl(\!\!\!\!&\bar{\hat m}{}^a\Bigl[
 2\bigl({}_0{\edth}{}_0{\edth}'\bar S\bigr)\bigl({}_0{\edth}t^{\bi}
 \bigr)+{}_0{\edth}\bigl(\bar S-S\bigr)t^{\bi}+\bigl(S+\bar S\bigr)
 \bigl({}_0{\edth}t^{\bi}\bigr)\Bigr]-\nonumber \\
\!\!\!\!&-\!\!\!\!&{\hat m}{}^a\Bigl[2\bigl({}_0{\edth}'{}_0{\edth}S
 \bigr)\bigl({}_0{\edth}'t^{\bi}\bigr)+{}_0{\edth}'\bigl(S-\bar S
 \bigr)t^{\bi}+\bigl(S+\bar S\bigr)\bigl({}_0{\edth}'t^{\bi}\bigr)
 \Bigr]\Bigr)R_{\bi}+\nonumber \\
+\frac{\rm i}{r}\!\!\!\!&\Bigl(\!\!\!\!&\bar{\hat m}{}^a{}_0{\edth}
 \bigl(S-\bar S\bigr)+{\hat m}{}^a{}_0{\edth}'\bigl(\bar S-S\bigr)
 \Bigr)R_0+ \nonumber \\
+\frac{2{\rm i}}{r}\!\!\!\!&\Bigl(\!\!\!\!&\bar{\hat m}{}^a\bigl({}
 _0{\edth}t^{\bi}\bigr)-\hat m{}^a\bigl({}_0{\edth}'t^{\bi}\bigr)
 \Bigr)G_{\bi}+O]\bigl(r^{-2}\bigr). \label{eq:5.3.9}
\end{eqnarray}
One can see that the coefficients of $R_{\bi}$ are real, but the 
coefficient of $R_0$ is {\em imaginary}. Consequently, in a general 
radiative spacetime (i.e. when the potential $S$ for the asymptotic 
shear {\em cannot} be chosen to be real) the reality of the whole 
$k^a$ {\em cannot} be ensured, even if $R_{\bi}$ and $G_{\bi}$ are 
chosen to be real. Indeed, the vanishing of $R_0$ is equivalent to 
$d=0$, which would imply $a=b=c=0$, too. 
Next recall that if $\xi^1:=-(1-\zeta^2)$, $\xi^2:=-{\rm i}(1+\zeta
^2)$ and $\xi^3:=-2\zeta$, then the general BMS vector field has the 
form 

\begin{eqnarray}
\hat K^a\!\!\!\!&=\!\!\!\!&\Bigl(H+\bigl(c_{\bi}+\bar c_{\bi}\bigr)
 t^{\bi}u\Bigr)\bigl(\frac{\partial}{\partial u}\bigr)^a+c_{\bi}\xi
 ^{\bi}\bigl(\frac{\partial}{\partial\zeta}\bigr)^a+\bar c_{\bi}\bar
 \xi^{\bi}\bigl(\frac{\partial}{\partial\bar\zeta}\bigr)^a= 
 \nonumber \\
\!\!\!\!&=\!\!\!\!&\Bigl(H+\bigl(c_{\bi}+\bar c_{\bi}\bigr)t^{\bi}
 u\Bigr)\bigl(\frac{\partial}{\partial u}\bigr)^a-2\bar c_{\bi}\bigl(
 {}_0{\edth}'t^{\bi}\bigr)\hat m^a-2c_{\bi}\bigl({}_0{\edth}t^{\bi}
 \bigr)\bar{\hat m}{}^a, \label{eq:5.3.9.bms}
\end{eqnarray}
where $H=H(\zeta,\bar\zeta)$ is an arbitrary real function and $c
_{\bi}$ are complex constants. This is a rotation BMS vector field 
tangent to the $u={\rm const}$ cut if $H=0$ and the constants $c
_{\bi}$ are purely imaginary $c_{\bi}={\rm i}R_{\bi}$. 
Therefore, taking the real part of $rk^a$, we have three independent 
real divergence free vector fields that are rotation BMS fields at 
the future null infinity, and, conversely, every rotation BMS vector 
field determines a vector field $rk^a$. Thus in the asymptotically 
flat context the role of the eigenvalue equation for the lowest 
eigenvalue is the unique extension of the BMS rotation vector fields 
off the future null infinity into a neighbourhood of the future null 
infinity. In addition, taking the imaginary part of $rk^a$ we have 
one divergence free vector field that is asymptotically vanishing as 
$1/r$. However, the parameterization of the latter is fixed by those 
of the real part of $rk^a$. All the ambiguities in $rk_a$ (including 
the addition of `gauge solutions') can be written as $rk_a\mapsto rk
_a+\frac{1}{r}G'_{\bi}K^{\bi}_a$ for some free (in general complex) 
{\em functions} $G'_{\bi}$ of $u$.

\subsubsection{The vector field $z^a_-$}
\label{sub-5.3.2}

Since the vector field $z^a_-$ is real, the function $F=r^2F^{(-2)}+
rF^{(-1)}+O(1)$, for which $rz^a_-=\varepsilon^{ab}\delta_bF$, can 
also be chosen to be real. Following the strategy of the previous 
subsection we can determine the explicit form of the components of 
$z^a_-$ in terms of the components of the spinor fields $\lambda_A$ 
and $\bar\mu_{A'}$, and comparing them with the defining equation for 
$F$, we obtain a system of differential equations for $F^{(-2)}$ and 
$F^{(-1)}$. Apart from the gauge terms, the solution will be an 
expression of 

\begin{equation}
\lambda^{(0)}_1\bigl({}_0{\edth}\bar\lambda^{(0)}_{0'}\bigr)-
\lambda^{(0)}_0\bigl({}_0{\edth}'\bar\lambda^{(0)}_{1'}\bigr)
=R_0+{\rm i}R_{\bi}t^{\bi}, \label{eq:5.3.10}
\end{equation}
where now the constants $R_0$ and $R_{\bi}$ are real and are given by 

\begin{eqnarray}
R_0\!\!\!\!&:=\!\!\!\!&-\frac{1}{4\sqrt{2}\pi}\Bigl({}_{(0)}c
 ^{-\frac{1}{2}}_1{}_{(0)}\bar c^{-\frac{1}{2}}_0+{}_{(0)}\bar c
 ^{\frac{1}{2}}_1{}_{(0)}c^{\frac{1}{2}}_0+{}_{(0)}\bar c
 ^{-\frac{1}{2}}_1{}_{(0)}c^{-\frac{1}{2}}_0+{}_{(0)}c^{\frac{1}{2}}
 _1{}_{(0)}\bar c^{\frac{1}{2}}_0\Bigr), \label{eq:5.3.11a} \\
R_1\!\!\!\!&:=\!\!\!\!&\frac{\rm i}{4\sqrt{2}\pi}\Bigl({}_{(0)}
 \bar c^{\frac{1}{2}}_1{}_{(0)}c^{-\frac{1}{2}}_0-{}_{(0)}c
 ^{-\frac{1}{2}}_1{}_{(0)}\bar c^{\frac{1}{2}}_0-{}_{(0)}c
 ^{\frac{1}{2}}_1{}_{(0)}\bar c^{-\frac{1}{2}}_0+{}_{(0)}\bar c
 ^{-\frac{1}{2}}_1{}_{(0)}c^{\frac{1}{2}}_0\Bigr), \label{eq:5.3.11b}\\
R_2\!\!\!\!&:=\!\!\!\!&\frac{1}{4\sqrt{2}\pi}\Bigl({}_{(0)}
 \bar c^{\frac{1}{2}}_1{}_{(0)}c^{-\frac{1}{2}}_0-{}_{(0)}c
 ^{-\frac{1}{2}}_1{}_{(0)}\bar c^{\frac{1}{2}}_0+{}_{(0)}c
 ^{\frac{1}{2}}_1{}_{(0)}\bar c^{-\frac{1}{2}}_0-{}_{(0)}\bar c
 ^{-\frac{1}{2}}_1{}_{(0)}c^{\frac{1}{2}}_0\Bigr), \label{eq:5.3.11c}\\
R_3\!\!\!\!&:=\!\!\!\!&\frac{\rm i}{4\sqrt{2}\pi}\Bigl({}_{(0)}c
 ^{-\frac{1}{2}}_1{}_{(0)}\bar c^{-\frac{1}{2}}_0+{}_{(0)}\bar c
 ^{\frac{1}{2}}_1{}_{(0)}c^{\frac{1}{2}}_0-{}_{(0)}\bar c
 ^{-\frac{1}{2}}_1{}_{(0)}c^{-\frac{1}{2}}_0-{}_{(0)}c^{\frac{1}{2}}
 _1{}_{(0)}\bar c^{\frac{1}{2}}_0\Bigr). \label{eq:5.3.11d}
\end{eqnarray}
Note that $R_0$ is independent of the $R_{\bi}$, in contrast to 
$R_0$ and $R_{\bi}$ of (\ref{eq:5.3.6a})-(\ref{eq:5.3.6d}). In 
terms of (\ref{eq:5.3.10}) the functions $F^{(-2)}$ and $F^{(-1)}$ 
take the simple, explicit form 

\begin{eqnarray}
F^{(-2)}=\!\!\!\!&-\!\!\!\!&2t^{\bi}R_{\bi}, \label{eq:5.3.12a} \\
F^{(-1)}=\!\!\!\!&-\!\!\!\!&2\Bigl(\bigl({}_0{\edth}S\bigr)\bigl({}_0
 {\edth}'t^{\bi}\bigr)+\bigl({}_0{\edth}'\bar S\bigr)\bigl({}_0
 {\edth}t^{\bi}\bigr)+\frac{1}{2}\bigl(S+\bar S\bigr)t^{\bi}\Bigr)
 R_{\bi}+ \nonumber \\
\!\!\!\!&+\!\!\!\!&{\rm i}\bigl(S-\bar S\bigr)R_0-2t^{\bi}G_{\bi},
 \label{eq:5.3.12b}
\end{eqnarray}
where the last term represents the ambiguity coming from the gauge 
solutions in the eigenspinor components $\lambda^{(1)}_{\uA}$; and, 
for the sake of simplicity, the irrelevant constants of integration 
have been chosen to be zero. Comparing (\ref{eq:5.3.12a}), 
(\ref{eq:5.3.12b}) with (\ref{eq:5.3.8a}), (\ref{eq:5.3.8b}) we see 
that, apart from constants and the $\pm$ sign, the coefficients of 
$R_{\bi}$ coincide, and the imaginary part of (\ref{eq:5.3.8b}) (i.e. 
the coefficient of $R_0$) is just the coefficient of the independent 
parameter $R_0$ in (\ref{eq:5.3.12b}). Therefore, the corresponding 
vector fields coincide: the vector fields $rz^a_-$ parameterized by 
$R_{\bi}$ are just the real part of the vector fields $rk^a$, and 
the asymptotically vanishing vector field, parameterized by $R_0$, 
is just the imaginary part of $rk^a$ of the previous subsection.

\subsection{2-surface observables at $\mathscr{I}^+$}
\label{sub-5.4}

\subsubsection{The observable $O[N^a]$}
\label{sub-5.4.1}

In \cite{Sz06} we showed that (1) the basic Hamiltonian of vacuum 
general relativity on a compact 3-manifold $\Sigma$ with smooth 
2-boundary ${\cal S}$ is functionally differentiable with respect to 
the ADM canonical variables if the area 2-form is fixed on ${\cal S}$, 
the lapse function is vanishing on ${\cal S}$, and the shift vector 
$N^a$ is tangent to ${\cal S}$ and divergence free with respect to 
the connection $\delta_e$; (2) the evolution equations preserve these 
boundary conditions; (3) the basic Hamiltonians form a closed Poisson 
algebra ${\cal H}_0$, in which the constraints form an ideal ${\cal 
C}$; (4) the value of the basic Hamiltonian on the constraint surface, 
given explicitly by the integral 

\begin{equation}
O\bigl[N^a\bigr]:=-\frac{1}{8\pi G}\oint_{\cal S}N^eA_e{\rm d}{\cal S},
\label{eq:5.4.1} 
\end{equation}
is a well defined, 2+2 covariant, gauge invariant observable; and (5) 
this $O$ provides a Lie algebra (anti-)homomorphism of the Lie algebra 
of the $\delta_e$-divergence-free vector fields on ${\cal S}$ into the 
quotient Lie algebra ${\cal H}_0/{\cal C}$ of observables. However, it 
should be noted that, independently of the quasi-local canonical 
analysis above, (\ref{eq:5.4.1}) already had appeared as a suggestion 
for the angular momentum of black hole horizons \cite{AsKr,BoFa,Gour}. 

We also considered the limit of this observable when the 2-surface 
${\cal S}$ tends to the spatial or the future null infinity. We showed 
that from the requirement of the finiteness of the limit at spatial 
infinity it follows that $N^a$ is necessarily a combination of the 
asymptotic rotation Killing vectors, and hence the corresponding 
observable is just the familiar expression of spatial angular momentum 
there. Unfortunately, however, at null infinity the general expression 
still contains a huge ambiguity. In fact, if the function $\nu$ 
defined by $N^a=\varepsilon^{ab}\delta_b\nu$ is written as $\nu=r^2
\nu^{(-2)}+r\nu^{(-1)}+O(1)$, then the $r\rightarrow\infty$ limit of 
$O[N^a]$ is finite precisely when $\nu^{(-2)}$ is a linear combination 
of the first four (i.e. $j=0$ and 1) ordinary spherical harmonics. In 
this case $N^a$ tends to a rotation BMS vector field, and for the 
observable $O[N^a]$, associated with a large sphere ${\cal S}_r$ of 
radius $r$, we obtain 

\begin{eqnarray}
O\bigl[N^a\bigr]=\frac{\rm i}{8\pi G}\oint\!\!\!\!\!\!&\Bigl(
 \!\!\!\!\!\!&\bigl(\psi^0_1+\sigma^0{}_0{\edth}\bar\sigma^0\bigr)
 \bigl({}_0{\edth}'\nu^{(-2)}\bigr)-\bigl(\bar\psi^0_1+\bar
 \sigma^0{}_0{\edth}'\sigma^0\bigr)\bigl({}_0{\edth}\nu^{(-2)}\bigr)+ 
 \nonumber \\
\!\!\!\!&+\!\!\!\!&\sigma^0\bigl({}_0{\edth}'{}^2\nu^{(-1)}\bigr)-
 \bar\sigma^0\bigl({}_0{\edth}{}^2\nu^{(-1)}\bigr)\Bigr){\rm d}{\cal 
 S}_1+O\bigl(r^{-1}\bigr), \label{eq:5.4.2} 
\end{eqnarray}
where ${\rm d}{\cal S}_1$ is the area element on the {\em unit} sphere. 
(In \cite{Sz06} $\nu^{(-2)}$ and $\nu^{(-1)}$ were denoted by $\nu
^{(2)}$ and $\nu^{(1)}$, respectively.) Though this formula reduces to 
the standard expression of angular momentum for stationary systems 
(when the potential $S$ for the asymptotic shear can be chosen to 
be real), in general spacetimes this is still ambiguous unless the 
function $\nu^{(-1)}$ has been specified. Or, in other words, the 
limit of $O[N^a]$ at the null infinity depends also on the way how the 
vector field $N^a$ tends to the rotation BMS vector field leaving the 
cut in question fixed. Thus to have a well defined angular momentum 
expression in a general, radiative spacetime within the framework 
defined by the observable $O[N^a]$, a prescription for $\nu^{(-1)}$ 
is needed.

\subsubsection{Spectral angular momentum and a measure of the magnetic 
part of the asymptotic shear at $\mathscr{I}^+$}
\label{sub-5.4.2}

The aim of the present subsection is to use the asymptotic rotation 
Killing fields $rk^a$ and $rz^a_-$ as the vector field $N^a$ and to 
calculate the $r\rightarrow\infty$ limit of the corresponding 
observable, hoping to obtain a reasonable definition of spatial 
angular momentum even in the presence of outgoing gravitational 
radiation. One way of doing this might be based on the observation 
that $N^aA_a=N^am_a(\bar\beta-\beta')+N^a\bar m_a(\beta-\bar\beta')=
N_{01'}\bar\tau+N_{10'}\tau$. Substituting the components of the 
vector field and the asymptotic form of the spin coefficient $\tau$ 
here, a rather lengthy calculation gives the desired expression. 

However, a much more economic method (and we adopt it here) is to use 
(\ref{eq:5.4.2}) by writing $\nu=F$ if $N^a$ is $rk^a$ or $rz^a_-$. 
Since only the {\em second} ${}_0{\edth}$ and ${}_0{\edth}'$ 
derivatives of $\nu^{(-1)}$ appear in (\ref{eq:5.4.2}), all the 
ambiguities are canceled from $\lim_{r\rightarrow\infty}O[N^a]$, i.e. 
{\em the observable $\lim_{r\rightarrow\infty} O[N^a]$ is well 
defined and does not depend on the gauge ambiguities of the solution 
$\lambda^{(1)}_{\uA}$}. Thus, for the sake of simplicity, the 
ambiguous terms in (\ref{eq:5.3.8b}) and (\ref{eq:5.3.12b}) can be 
chosen to be zero: $G_0=0$ and $G_{\bi}=0$. 

To determine the explicit form of the observable $\lim_{r\rightarrow
\infty} O[N^a]$, recall that the functions $\nu^{(-2)}$ and $\nu
^{(-1)}$ are linear functions of the parameters $R_{\bi}$ and $R_0$. 
Therefore, the $r\rightarrow\infty$ limit of the observables $O[rk^a]$ 
and $O[rz^a_-]$ have the structure 

\begin{eqnarray}
\pm\frac{1}{2}\lim_{r\rightarrow\infty}O\bigl[rk^a\bigr]\!\!\!\!&=
 :\!\!\!\!&R_{\bi}{\tt J}^{\bi}+{\rm i}R_0{\tt M}, \label{eq:5.4.3a} \\
\frac{1}{2}\lim_{r\rightarrow\infty}O\bigl[rz^a_-\bigr]\!\!\!\!&=
 :\!\!\!\!&R_{\bi}{\tt J}^{\bi}+R_0{\tt M}, \label{eq:5.4.3b} 
\end{eqnarray}
respectively. From the actual form of $F^{(-1)}$ it follows that 
${\tt J}^{\bi}$ and ${\tt M}$ are real. Recalling that the vector 
fields $rk^a$ and $rz^a_-$ tend (or can be chosen to tend) 
asymptotically to the combination $-2R_{\bi}K^{\bi}_a$ of the rotation 
Killing fields of the metric sphere of radius $r$ with {\em real} 
$R_{\bi}$, which have in fact unique extension to real BMS vector 
fields tangent to the $u={\rm const}$ cut of $\mathscr{I}^+$, the 
observable ${\tt J}^{\bi}$ may be interpreted as the spatial angular 
momentum associated with the $u={\rm const}$ cut of the future null 
infinity. Remarkably enough, {\em the vector fields $rk^a$ and $rz
^a_-$ define the same observables ${\tt J}^{\bi}$ and ${\tt M}$}, 
i.e. the ${\tt J}^{\bi}$'s and the ${\tt M}$'s in equations 
(\ref{eq:5.4.3a}) and (\ref{eq:5.4.3b}) {\em coincide}. 

The explicit form of ${\tt J}^{\bi}$ is 

\begin{eqnarray}
{\tt J}^{\bi}=\frac{\rm i}{8\pi G}\oint\!\!\!\!&\Bigl\{\!\!\!\!&
 \Bigl(\psi^0_1+2\sigma^0{}_0{\edth}\bar\sigma^0\Bigr)\bigl({}_0
 {\edth}'t^{\bi}\bigr)-\Bigl(\bar\psi^0_1+2\bar\sigma^0{}_0{\edth}'
 \sigma^0\Bigr)\bigl({}_0{\edth}t^{\bi}\bigr)+\nonumber \\
\!\!\!\!&+\!\!\!\!&\sigma^0\Bigl({}_0{\edth}\bigl({}_0{\edth}'^2S
 \bigr)\bigl({}_0{\edth}'t^{\bi}\bigr)+\frac{1}{2}\bigl({}_0
 {\edth}'^2S\bigr)t^{\bi}\Bigr)-\nonumber \\
\!\!\!\!&-\!\!\!\!&\bar\sigma^0\Bigl({}_0{\edth}'\bigl({}_0{\edth}^2
 \bar S\bigr)\bigl({}_0{\edth}t^{\bi}\bigr)+\frac{1}{2}\bigl({}_0
 {\edth}^2\bar S\bigr)t^{\bi}\Bigr)\Bigr\}{\rm d}{\cal S}_1. 
\label{eq:5.4.4}
\end{eqnarray}
It could be interesting to note that the first line gives just the 
spatial part of Bramson's angular momentum expression \cite{Br}: in 
fact, taking into account that ${}_0{\edth}(\sigma^0\bar\sigma{}^0)
({}_0{\edth}'t^{\bi})={}_0{\edth}(\sigma^0\bar\sigma{}^0{}_0{\edth}'t
^{\bi})+\sigma^0\bar\sigma{}^0t^{\bi}$, the first line of the 
integrand can be written as the imaginary part of $(\psi^0_1+2\sigma
^0{}_0{\edth}\bar\sigma^0+{}_0{\edth}(\sigma^0\bar\sigma{}^0))({}_0
{\edth}'t^{\bi})$. Therefore, the whole ${\tt J}^{\bi}$ can be 
interpreted as Bramson's angular momentum plus some `correction terms' 
built from the potential $S$ in a gauge invariant way. 

Apart from the Weyl spinor component $\psi^0_1$ all the terms of the 
integrand in (\ref{eq:5.4.4}) can be written into the form of a 
bilinear expression of the real and the imaginary parts of the 
potential $S$. In fact, a trivial calculation gives that the terms in 
the integrand of (\ref{eq:5.4.4}) containing $t^{\bi}$ algebraically 
can be written as 

$$
\frac{1}{4}\Bigl({}_0{\edth}^2\bigl(S+\bar S\bigr){}_0{\edth}'{}^2
\bigl(S-\bar S\bigr)+{}_0{\edth}'{}^2\bigl(S+\bar S\bigr){}_0{\edth}
^2\bigl(S-\bar S\bigr)\Bigr)t^{\bi};
$$
while the remaining terms containing the asymptotic shear as 

\begin{eqnarray}
\frac{1}{4}\Bigl\{\!\!\!\!&\Bigl(\!\!\!\!&3{}_0{\edth}^2\bigl(S+\bar 
 S\bigr){}_0{\edth}{}_0{\edth}'{}^2\bigl(S+\bar S\bigr)-
 {}_0{\edth}^2\bigl(S-\bar S\bigr){}_0{\edth}{}_0{\edth}'{}^2\bigl(
 S-\bar S\bigr)\Bigr)\bigl({}_0{\edth}'t^{\bi}\bigr)- \nonumber \\
-\!\!\!\!&\Bigl(\!\!\!\!&3{}_0{\edth}'{}^2\bigl(S+\bar S\bigr){}_0
 {\edth}'{}_0{\edth}^2\bigl(S+\bar S\bigr)-{}_0{\edth}'{}^2\bigl(S-
 \bar S\bigr){}_0{\edth}'{}_0{\edth}^2\bigl(S-\bar S\bigr)\Bigr)
 \bigl({}_0{\edth}t^{\bi}\bigr)\Bigr\}+ \nonumber \\
+\frac{1}{4}\Bigl\{\!\!\!\!&\Bigl(\!\!\!\!&3{}_0{\edth}^2\bigl(S-
 \bar S\bigr){}_0{\edth}{}_0{\edth}'{}^2\bigl(S+\bar S\bigr)-
 {}_0{\edth}^2\bigl(S+\bar S\bigr){}_0{\edth}{}_0{\edth}'{}^2\bigl(
 S-\bar S\bigr)\Bigr)\bigl({}_0{\edth}'t^{\bi}\bigr)+ \nonumber \\
+\!\!\!\!&\Bigl(\!\!\!\!&3{}_0{\edth}'{}^2\bigl(S-\bar S\bigr){}_0
 {\edth}'{}_0{\edth}^2\bigl(S+\bar S\bigr)-{}_0{\edth}'{}^2\bigl(S+
 \bar S\bigr){}_0{\edth}'{}_0{\edth}^2\bigl(S-\bar S\bigr)\Bigr)
 \bigl({}_0{\edth}t^{\bi}\bigr)\Bigr\}. \label{eq:5.4.4a}
\end{eqnarray}
However, the first two lines of (\ref{eq:5.4.4a}), i.e. the terms 
quadratic in $(S+\bar S)$ and in $(S-\bar S)$, can be written into 
a total divergence. Indeed, by taking total derivatives and using 
the commutator of the edth operators, for any function $f$ we obtain 

\begin{eqnarray}
\bigl({}_0{\edth}^2f\bigr){}_0{\edth}\bigl({}_0{\edth}'{}^2f\bigr)
 \bigl({}_0{\edth}'t^{\bi}\bigr)\!\!\!\!&=\!\!\!\!&{}_0{\edth}\Bigl(
 \bigl({}_0{\edth}f\bigr){}_0{\edth}\bigl({}_0{\edth}'{}^2f\bigr)
 \bigl({}_0{\edth}'t^{\bi}\bigr)+\frac{1}{2}\bigl({}_0{\edth}{}_0
 {\edth}'f\bigr)^2\bigl({}_0{\edth}'t^{\bi}\bigr)\Bigr)+ \nonumber \\
\!\!\!\!&+\!\!\!\!&{}_0{\edth}'\Bigl(\bigl({}_0{\edth}f\bigr)\bigl(
 {}_0{\edth}{}_0{\edth}'f\bigr)t^{\bi}-\bigl({}_0{\edth}f\bigr){}_0
 {\edth}'\bigl({}_0{\edth}^2f\bigr)\bigl({}_0{\edth}'t^{\bi}\bigr)
 \Bigr)+\nonumber \\ 
\!\!\!\!&+\!\!\!\!&\Bigl(\bigl({}_0{\edth}f\bigr)\bigl({}_0{\edth}'
 f\bigr)-\frac{1}{2}\bigl({}_0{\edth}{}_0{\edth}'f\bigr)^2\Bigr)
 t^{\bi},\nonumber 
\end{eqnarray}
implying that $({}_0{\edth}^2f){}_0{\edth}({}_0{\edth}'{}^2f)({}_0
{\edth}'t^{\bi})-({}_0{\edth}'{}^2f){}_0{\edth}'({}_0{\edth}^2f)({}_0
{\edth}t^{\bi})$ is a total divergence. Applying this to $f=S+\bar S$ 
and to $f=S-\bar S$ in (\ref{eq:5.4.4a}) we finally obtain that the 
first two lines in (\ref{eq:5.4.4a}) do, indeed, form a total 
divergence. This implies, in particular, that in stationary spacetimes, 
when $S$ is real (see \cite{NP}), ${\tt J}^{\bi}$ reduces to the 
standard expression \cite{Br78} 

$$
\frac{\rm i}{8\pi G}\oint\Bigl(\psi^0_1\bigl({}_0{\edth}'t^{\bi}
\bigr)-\bar\psi^0_1\bigl({}_0{\edth}t^{\bi}\bigr)\Bigr){\rm d}S_1.
$$
We also note that in axi-symmetric spacetime with Killing vector $K^a$ 
the observable $O[K^a]$ on axi-symmetric surfaces is just the Komar 
expression \cite{Sz06}. 
We call the ${\tt J}^{\bi}$ given by (\ref{eq:5.4.4}) the {\em spectral 
angular momentum} at $\mathscr{I}^+$. 

The explicit form of the other observable ${\tt M}$ is 

\begin{equation}
{\tt M}=\frac{1}{16\pi G}\oint\Bigl(2\sigma^0\bar\sigma^0-\bigl(
{}_0{\edth}'^2\bar S\bigr)\bigl({}_0{\edth}^2\bar S\bigr)-\bigl(
{}_0{\edth}'^2S\bigr)\bigl({}_0{\edth}^2S\bigr)\Bigr){\rm d}{\cal 
S}_1. \label{eq:5.4.5}
\end{equation}
Clearly, this is real and is vanishing for purely electric $\sigma
^0$, i.e. when $S$ is real. However, this is, in fact, {\em 
non-negative and zero precisely for purely electric asymptotic 
shear}; i.e. it defines a measure of the presence of the magnetic 
part of the asymptotic shear. To see this, let us rewrite its 
integrand (by integration by parts and by using ${}_0{\edth}^2{}_0
{\edth}'{}^2S={}_0{\edth}'{}^2{}_0{\edth}^2S$) as 

\begin{eqnarray}
\!\!\!\!&{}\!\!\!\!&\bigl({}_0{\edth}^2S\bigr){}_0{\edth}'{}^2\bigl(
 \bar S-S\bigr)+\bigl({}_0{\edth}'{}^2\bar S\bigr){}_0{\edth}^2
 \bigl(S-\bar S\bigr)\simeq \nonumber \\
\!\!\!\!&\simeq\!\!\!\!&\bigl({}_0{\edth}'{}^2{}_0{\edth}^2S\bigr)
 \bigl(\bar S-S\bigr)+\bigl({}_0{\edth}'{}^2\bar S\bigr){}_0{\edth}^2
 \bigl(S-\bar S\bigr)= \nonumber \\
\!\!\!\!&=\!\!\!\!&\bigl({}_0{\edth}^2{}_0{\edth}'{}^2S\bigr)
 \bigl(\bar S-S\bigr)+\bigl({}_0{\edth}'{}^2\bar S\bigr){}_0{\edth}^2
 \bigl(S-\bar S\bigr)\simeq \nonumber \\
\!\!\!\!&\simeq\!\!\!\!&\Bigl({}_0{\edth}'{}^2\bigl(\bar S-S\bigr)
 \Bigr)\Bigl({}_0{\edth}^2\bigl(S-\bar S\bigr)\Bigr),\nonumber
\end{eqnarray}
where $\simeq$ means `equal up to total divergences'. Since in 
stationary spacetimes (and hence in the absence of outgoing 
gravitational radiation) the asymptotic shear is always purely 
electric \cite{NP}, the non-vanishing of ${\tt M}$ indicates 
non-trivial dynamics of the gravitational field near the future null 
infinity. 

From (\ref{eq:5.4.4}) and (\ref{eq:5.4.5}) it is clear that both 
${\tt J}^{\bi}$ and ${\tt M}$ are well defined in the sense that they 
depend only on the cut of $\mathscr{I}^+$. In particular, they do 
{\em not} depend on the choice of the origin of $u$, and hence ${\tt 
J}^{\bi}$ is free of supertranslation ambiguities. The notion of 
spectral angular momentum is based on the solution of a certain 
elliptic equation on the cut rather than on the BMS vector fields. 
The latter is used only to {\em interpret} ${\tt J}^{\bi}$ as spatial 
angular momentum. On the other hand, since the vector field $N^e$ in 
the observable $O[N^e]$ must be tangent to the 2-surface, in the 
theoretical framework based on $O[N^e]$ we cannot ask for the effect 
of (any form of) translations or boosts. Thus the spectral angular 
momentum ${\tt J}^{\bi}$ should probably be interpreted only as the 
{\em spin part} of the (as yet not known) total {\em relativistic} 
angular momentum.

\subsubsection{Comparison of angular momenta on different cuts} 
\label{sub-5.4.3}

In the previous subsection, we saw that ${\tt J}^{\bi}$ is 
unambiguously associated with any cut of $\mathscr{I}^+$. In the 
present subsection, we ask how the angular momenta associated with 
two {\em different} cuts of $\mathscr{I}^+$, say $\hat{\cal S}$ and 
$\tilde{\hat{\cal S}}$, can be compared. First we discuss the 
theoretical basis of this comparison, and then we derive the formula 
for the flux of spectral angular momentum carried away by the 
outgoing gravitational radiation in Einstein--Maxwell spacetimes. 
(The hat over a symbol is referring to the unphysical spacetime, 
indicating that that may be ill-defined in the physical spacetime.) 

Mathematically ${\tt J}^{\bi}$ is an element of the dual of the 
space ${\cal R}$ of the rotation BMS vector fields that are tangent 
to $\hat{\cal S}$. Hence the spectral angular momenta on the cuts 
$\hat{\cal S}$ and $\tilde{\hat{\cal S}}$ can be compared only if we 
can find a natural isomorphism $I:{\cal R}\rightarrow\tilde{\cal R}$, 
yielding the identification of the corresponding dual spaces as well. 
Formally, this isomorphism is analogous to that we already discussed 
in a nutshell at the end of subsection \ref{sub-5.2} in a slightly 
different context. 
Thus, let us fix the Bondi conformal gauge on $\mathscr{I}^+$, and 
hence both $\hat{\cal S}$ and $\tilde{\hat{\cal S}}$ inherit the 
unit sphere metric, and let $\hat n^a$ be the future pointing tangent 
of the null geodesic generators of $\mathscr{I}^+$ such that $\hat n
_a=-\hat\nabla_a\Omega\vert_{\mathscr{I}^+}$. The spinor constituent 
$\hat\iota^A$ of $\hat n^A$ can be chosen to be constant along the 
generators: $\hat n^a\hat\nabla_a\hat\iota^B=0$, and fix its phase. 
Then the spinor $\hat\iota^A$ can be completed to a spin frame 
$\{\hat o^A,\hat\iota^A\}$ at the points of the cuts $\hat{\cal S}$ 
and $\tilde{\hat{\cal S}}$ such that the complex null vectors $\hat m
^a:=\hat o^A\bar{\hat\iota}{}^{A'}$ and $\bar{\hat m}{}^a:=\hat\iota
^A\bar{\hat o}{}^{A'}$ are tangent to the surfaces. This $\hat o^A$ 
is uniquely determined by the cuts, and hence we have a uniquely 
determined complex null basis $\{\hat m^a,\bar{\hat m}{}^a\}$ on both 
$\hat{\cal S}$ and $\tilde{\hat{\cal S}}$. (If the cuts intersect 
each other, then, of course, the spinors $\hat o^A$ at the points of 
intersection do not coincide unless the cuts are tangent to each 
other there.) In addition, if we fix a complex stereographic 
coordinate system $(\zeta,\bar\zeta)$ on one of the cuts, then by the 
condition $\hat n^a\hat\nabla_a\zeta=0$ this determines a complex 
stereographic coordinate system on the other cut (and, in fact, on 
the whole $\mathscr{I}^+$) in a unique way. 
But then, recalling the structure (\ref{eq:5.3.9.bms}) of the 
rotation BMS vector fields, we have an isomorphism $I:{\cal R}
\rightarrow\tilde{\cal R}$: the vector field $\hat K^a_0\in{\cal R}$ 
is identified with $\hat K^a_1\in\tilde{\cal R}$ via $I$ precisely 
when their components $c_{\bi}$ coincide. It is this isomorphism by 
means of which we identify the dual spaces ${\cal R}^*$ and $\tilde
{\cal R}^*$ with each other, and one can subtract ${\tt J}^{\bi}$ 
from $\tilde{\tt J}^{\bi}$ directly. Thus we are ready to calculate 
how ${\tt J}^{\bi}$ changes from cut to cut. 

Thus let $u$ be the Bondi time coordinate whose origin is chosen to 
be $\hat{\cal S}$, introduce the 1-parameter family of cuts $\hat
{\cal S}_u$ obtained from $\hat{\cal S}$ by BMS time translations 
along $\hat n^a$, denote the spectral angular momentum on $\hat{\cal 
S}_u$ by ${\tt J}^{\bi}(u)$, and calculate its derivative with 
respect to $u$ at $u=0$. If, for the sake of brevity, we write the 
integrand of ${\tt J}^{\bi}$ as $(\psi^0_1+F){}_0{\edth}'t^{\bi}-
(\bar\psi^0_1+\bar F){}_0{\edth}t^{\bi}$ (by using ${}_0{\edth}{}_0
{\edth}'t^{\bi}={}_0{\edth}'{}_0{\edth}t^{\bi}=-t^{\bi}$), then by 
(\ref{eq:5.0.1a}) and (\ref{eq:5.0.1c}) we find 

\begin{eqnarray}
\dot{\tt J}^{\bi}=\frac{\rm i}{8\pi G}\oint\Bigl(4\!\!\!\!&G\!\!\!\!&
 \bigl(\varphi^0_1\bar\varphi^0_2\bigl({}_0{\edth}'t^{\bi}\bigr)-
 \varphi^0_2\bar\varphi^0_1\bigl({}_0{\edth}t^{\bi}\bigr)\bigr)+ 
 \nonumber \\
+\!\!\!\!&t^{\bi}\!\!\!\!&\bigl(\bar\sigma^0\dot\sigma^0+2{}_0
 {\edth}'\bigl(\sigma^0{}_0{\edth}\dot{\bar\sigma}{}^0\bigr)-{}_0
 {\edth}'\dot F\bigr)- \nonumber \\
-\!\!\!\!&t^{\bi}\!\!\!\!&\bigl(\sigma^0\dot{\bar\sigma}{}^0+2{}_0
 {\edth}\bigl(\bar\sigma^0{}_0{\edth}'\dot\sigma^0\bigr)-{}_0{\edth}
 \dot{\bar F}\bigr)\Bigr){\rm d}{\cal S}_1.    
\label{eq:5.4.6}
\end{eqnarray}
The first line is just the angular momentum flux carried away by the 
electromagnetic radiation, while the remaining terms can be 
interpreted as that carried away by the gravitational waves. In fact, 
the first two terms can also be written as 

$$
\Phi^{\bi}:=\frac{1}{8\pi}\lim_{r\rightarrow\infty}\oint_{{\cal S}_r}
n^aT_{ab}2{\rm i}\Bigl(\hat m^b\bigl({}_0{\edth}'t^{\bi}\bigr)-\bar
{\hat m}{}^b\bigl({}_0{\edth}t^{\bi}\bigr)\Bigr){\rm d}{\cal S}_r,
$$
where $T_{ab}=2\varphi_{AB}\bar\varphi_{A'B'}$, the energy-momentum 
tensor of the Maxwell field. Recalling the form (\ref{eq:5.3.9.bms}) 
of the rotation BMS vector field, $\Phi^{\bi}du$ is just the angular 
momentum current carried away by the electromagnetic radiation 
between the $u$ and $u+du$ retarded times. 

Moreover, recall that by (\ref{eq:5.0.1b}) the time derivative of 
the Bondi--Sachs energy-momentum, ${\tt P}^{\ua}:=-\frac{1}{4\pi G}
\oint(\psi^0_2+\sigma^0\dot{\bar\sigma}{}^0)t^{\ua}{\rm d}{\cal S}_1$, 
is the integral of $-\frac{1}{4\pi G}(\dot\sigma^0\dot{\bar\sigma}{}
^0+2G\varphi^0_2\bar\varphi^0_2)t^{\ua}$, where $t^{\ua}:=(1,t
^{\bi})$ (see e.g. \cite{NP,NT,PRII}). Thus the vanishing of the 
outgoing energy flux is equivalent to $\dot\sigma^0=0$ and $\varphi
^0_2=0$. Clearly, if $\dot\sigma^0=0$, then the corresponding shear 
potential is also time independent: $\dot S=0$. Then, however, 
equation (\ref{eq:5.4.6}) shows, in particular, that {\em in the 
absence of outgoing energy flux the outgoing spectral angular 
momentum flux is also vanishing}. Obviously, every angular momentum 
expression shares this property which can be written as the integral 
of the imaginary part of $(\psi^0_1+F){}_0{\edth}'t^{\bi}$ with some 
gauge invariant expression $F$ of $S$ and $\bar S$. 

Although we could consider a more general family of cuts obtained from 
$\hat{\cal S}$ by a general {\em supertranslation} along $H\hat n^a$ 
with a general $H=H(\zeta,\bar\zeta)$, but the resulting formula does 
not seem to yield much deeper understanding of gravitational radiation. 
Technically, the use of such a more general foliation of $\mathscr{I}
^+$ yields only that the $\hat o^A$ spinor of the spin frame adapted 
to the foliation undergoes the special null rotation $\hat o^A\mapsto
\hat o^A+({}_0{\edth}H)\hat\iota^A$, and the $u$-derivatives $\dot
\psi^0_1$ and $\dot\sigma^0$ have to be substituted by $H\dot\psi^0_1
+3({}_0{\edth}H)\psi^0_2$ and $H\dot\sigma^0-{}_0{\edth}^2H$, 
respectively. 

The ultimate answer whether or not the spectral angular momentum and 
the observable ${\tt M}$ have physical significance will be given by 
the practice. In particular, it could be interesting to see whether 
or not the inequalities like in \cite{CT,Da} can be proven for ${\tt 
J}^{\bi}$ too, or whether in the unified model of spatial and null 
infinity \cite{HF} the $u\rightarrow-\infty$ limit of ${\tt J}^{\bi}$ 
reduces to the ADM angular momentum.

\section{Appendix}
\label{sec-6}

\subsection{Hermitian fibre metrics on the spinor bundle}
\label{sub-6.1}

Clearly, any Hermitian fibre metric $G_{AA'}$ on ${\mathbb S}^A({\cal 
S})$ can also be considered as a {\em real} section $G_a$ of the 
(dual) Lorentzian vector bundle ${\mathbb V}_a({\cal S})$, defined 
to be the pull back to ${\cal S}$ of the spacetime cotangent bundle 
$T^*M$. This $G_{AA'}$ is positive definite (and hence nonsingular) 
iff it is future pointing and timelike in ${\mathbb V}_a({\cal S})$. 
Such a metric can be specified by four real functions on ${\cal S}$. 

Next introduce the inverse $G^{AA'}$ of $G_{AA'}$ by $G^{AA'}
G_{BA'}=\delta^A_B$. Then from $(\varepsilon^{AB}\varepsilon^{A'B'}
G_{BB'})G_{CA'}=\frac{1}{2}\delta^A_CG_eG_fg^{ef}$ we see that the 
inverse $G^{AA'}$ is just the contravariant form of $G_{AA'}$, i.e. 
$G^{AA'}=\varepsilon^{AB}\varepsilon^{A'B'}G_{BB'}$ (or, 
equivalently, the symplectic and the Hermitian metrics are 
compatible in the sense that $\varepsilon^{AB}G_{AA'}G_{BB'}=
\varepsilon_{A'B'}$) iff $G_aG_bg^{ab}=2$. This normalization 
reduces the independent components of $G_{AA'}$ to three. However, 
such a fibre metric can be specialized further by requiring its 
compatibility with the chirality: $G_{AA'}\gamma^A{}_B\bar\gamma
^{A'}{}_{B'}$ $=G_{BB'}$. Since $\gamma^A{}_B\bar\gamma^{A'}{}_{B'}$, 
as a base-point preserving bundle map ${\mathbb V}_a({\cal S})
\rightarrow{\mathbb V}_a({\cal S})$ acts as identity precisely on 
sections orthogonal to ${\cal S}$ (and as minus the identity 
precisely on sections tangent to ${\cal S}$), this compatibility 
is equivalent to the orthogonality of $G_a$ to ${\cal S}$. The 
independent components of such a metric is only one. In a fixed, 
normalized GHP spin frame $\{o^A,\iota^A\}$, adapted to ${\cal S}$, 
it has the general form $G_{AA'}=go_A\bar o_{A'}+(1/g)\iota_A\bar
\iota_{A'}$, where $g$ is an arbitrary, strictly positive real 
function on ${\cal S}$. Note that for orientable ${\cal S}$ in a 
time and space orientable spacetime there is no obstruction to the 
global existence of such a $G_{AA'}$, because in this case the normal 
bundle of ${\cal S}$ in the spacetime is globally trivializable. 

On general 2-surfaces there does not seem to be any natural choice 
for such a $G_{AA'}$, or, equivalently, for such a $g$. If, however, 
the 2-surface is mean-convex, i.e. when the dual mean curvature 
vector of ${\cal S}$ is timelike, then there is such a choice. In 
fact, the mean curvature vector $Q_b:=Q^a_{ab}=-2(\rho'o_A\bar o_{A'}
+\rho\iota_A\bar\iota_{A'})$ and its dual, $\tilde Q_b:={}^\bot
\varepsilon_{ba}Q^a=2(\rho'o_A\bar o_{A'}-\rho\iota_A\bar\iota
_{A'})$, are globally defined and orthogonal to ${\cal S}$, 
furthermore $\vert Q\vert^2:=\tilde Q_a\tilde Q^a=-Q_aQ^a=-8\rho
\rho'$. Then for mean convex surfaces either $\rho>0$ and $\rho'<0$ 
or $\rho<0$ and $\rho'>0$, and it is natural to define $G_{AA'}:=
\pm\sqrt{2}\tilde Q_{AA'}/\vert \tilde Q\vert=\pm(\rho' o_A\bar 
o_{A'}-\rho\iota_A\bar\iota_{A'})/\sqrt{\vert\rho\rho'\vert}$ with 
the sign yielding future pointing $G^a$.

\subsection{Non-existence of constant Hermitian fibre metrics}
\label{sub-6.2}

We show that the existence of a positive definite Hermitian metric on 
the spinor bundle compatible with the covariant derivative operator 
$\delta_e$ is equivalent to the triviality of the holonomy of $\delta
_e$ on the normal bundle. 

Let $G_{AA'}$ be a Hermitian scalar product on the spinor bundle 
${\mathbb S}^A({\cal S})$ and assume that $\delta_eG_{AA'}=0$. 
Considering this metric to be a Lorentzian vector field on ${\cal 
S}$, the integrability condition of $\delta_eG_{AA'}=0$ is $0=
(\delta_c\delta_d-\delta_d\delta_c)G_b=G_af^a{}_{bcd}$. Contracting 
this with the area 2-form $\varepsilon^{cd}$ and taking into 
account the explicit expression for the curvature $f^a{}_{bcd}$ 
given in subsection \ref{sub-2.1}, we obtain 

\begin{equation}
\Bigl(R\varepsilon_{ab}-2\varepsilon^{cd}\bigl(\delta_cA_d\bigr)
{}^\bot\varepsilon_{ab}\Bigr)G^b=0.\label{eq:A.1}
\end{equation}
This implies that $\delta_e$ admits a constant section of the vector 
bundle ${\mathbb V}_a({\cal S})$ only if $\delta_e$ is flat. 
Restricting $G_a$ to be normal to ${\cal S}$ we find in particular 
that $\delta_e$ admits a constant section of the normal bundle only 
if $A_e$ is flat. (By (\ref{eq:A.1}) $\delta_e$ admits a constant 
section of the tangent bundle only if $({\cal S},q_{ab})$ is flat, 
which, by the Gauss--Bonnet theorem, can happen only if ${\cal S}$ 
is a torus.) Therefore, the necessary condition of the existence of 
a $\delta_e$-constant positive definite Hermitian metric, being 
compatible with $\gamma^A{}_B$, is the flatness of the connection 
$\delta_e$ on the normal bundle. Then, however, the globality of 
$G_a$ implies that the connection is holonomically trivial too. To 
see this, consider a closed curve $\gamma:[0,1]\rightarrow{\cal S}$ 
with the base point $p:=\gamma(0)=\gamma(1)$, and suppose, on the 
contrary, that the holonomy $H_\gamma:N_p{\cal S}\rightarrow N_p
{\cal S}$, as an element of the structure group $SO(1,1)$, is 
different from identity. Since $G^a$ is globally defined on ${\cal 
S}$ and constant with respect to $\delta_e$, the holonomy $H
_\gamma$ acts on $G^a_p$ (the value of $G^a$ at $p$) as the 
identity. Since the only element of $SO(1,1)$ leaving the nonzero 
vector $G^a$ fixed is the identity, the holonomy $H_\gamma$ is the 
identity, and hence the whole holonomy group at every point $p$, 
must be trivial. 

Conversely, the triviality of the holonomy of the connection 
$\delta_e$ on the normal bundle and the global trivializability of 
$N{\cal S}$ imply the existence of a globally defined orthonormal 
$\delta_e$-constant frame field $\{t^a,v^a\}$ with future pointing 
and timelike $t^a$. Then e.g. $G_{AA'}:=\sqrt{2}t_{AA'}$ is a desired 
positive definite Hermitian scalar product on the spinor bundle. 

\hskip 25pt

The author is grateful to J\"org Frauendiener and Ted Newman for the 
useful discussions on the angular momentum at null infinity. This 
work was partially supported by the Hungarian Scientific Research 
Fund (OTKA) grants T042531 and K67790.


\noindent

\begin{thebibliography}{9999}

\bibitem{Win} J. Winicour, Angular momentum in general relativity, in 
         {\it General Relativity and Gravitation}, Vol 2, pp 71--96, 
         Ed. A. Held, Plenum Publishing Co, New York 1980

\bibitem{Sz01}
         L. B. Szabados, On certain quasi-local spin-angular momentum 
          expressions for large spheres near the null infinity, Class. 
          Quantum Grav. {\bf 18}  5487--5510 (2001), gr-qc/0109047

\bibitem{Mo} 
         O. M. Moreschi, Intrinsic angular momentum and centre of 
         mass in general relativity, Class. Quantum Grav. {\bf 21} 
         5409--5425 (2004), gr-qc/0209097

\bibitem{CJK} 
         P. T. Chru\'sciel, J. Jezierski, J. Kijowski, {\it A 
         Hamiltonian Framework for Field Theories in the Radiative 
         Regime}, Vol m70 of Lecture Notes in Physics, Springer, 
         Berlin, Heidelberg, New York 2001

\bibitem{Sz04}
         L. B. Szabados, Quasi-local energy-momentum and angular 
          momentum in GR: A review article, Living Rev. Relativity 
          {\bf 7} (2004) 4

\bibitem{PRII}
         R. Penrose, W. Rindler, {\it Spinors and Spacetime}, Vol 2, 
          Cambridge University Press, Cambridge 1986

\bibitem{He} A. D. Helfer, Angular momentum of isolated systems, 
         Gen. Rel. Grav. (to be published), arXiv:0709.1078

\bibitem{Sz06} 
         L. B. Szabados, On a class of 2-surface observables in general 
          relativity, Class. Quantum Grav. {\bf 23} 2291--2302 (2006), 
          gr-qc/0511059

\bibitem{CW} G. B. Cook, B. F. Whiting, Approximate Killing vectors on 
        $S^2$, Phys. Rev. D {\bf 76} 041501(R) (2007), arXiv:0706.0199

\bibitem{ECS}
         M. Engman, R. Cordero-Soto, Intrinsic spectral geometry of the 
         Kerr--Newman event horizon, J. Math. Phys. {\bf 47} 033503-1--6 
         (2005), math-ph/0509067

\bibitem{TFr} Th. Friedrich, Der erste Eigenwert des Dirac-operators 
         einer kompakten Riemannschen Mannigfaltigkeit nichtnegativer 
         Skalarkr\"ummung, Math. Nachr. {\bf 97} 117--146 (1980)

\bibitem{Hi86} O. Hijazi, A conformal lower bound for the smallest 
         eigenvalue of the Dirac operator and Killing spinors, Commun. 
         Math. Phys. {\bf 104} 151--162 (1986)

\bibitem{Hi95} O. Hijazi, Lower bounds for the eigenvalues of the 
         Dirac operator, J. Geom. Phys. {\bf 16} 27--38 (1995) 

\bibitem{TF00} Th. Friedrich, {\it Dirac Operators in Riemannian 
         Geometry}, Graduate Studies in Mathematics, Vol 25, AMS 
         Providence, Rhode Island 2000

\bibitem{Ba92} C. B\"ar, Lower eigenvalue estimates for Dirac 
         operators, Math. Ann. {\bf 293} 39--46 (1992) 

\bibitem{FK} Th. Friedrich, E. C. Kim, Some remarks on the Hijazi 
         inequality and generalizations of the Killing equation for 
         spinors, J. Geom. Phys. {\bf 37} 1--14 (2001)

\bibitem{Zh} X. Zhang, Lower bounds for eigenvalues of hypersurface 
          Dirac operators, Math. Res. Lett. {\bf 5} 199--210 (1998)


\bibitem{Sz94} 
         L. B. Szabados, Two dimensional Sen connections in general 
          relativity, Class. Quantum Grav. {\bf 11} 1833--1846 (1994),
          gr-qc/9402001

\bibitem{L} A. Lichnerowicz, Spineurs harmoniques, C. R. Acad. Sci. 
          Paris A--B {\bf 257} 7--9 (1963)

\bibitem{PRI}
         R. Penrose, W. Rindler, {\it Spinors and Spacetime}, Vol 1, 
          Cambridge University Press, Cambridge 1984

\bibitem{GHP} R. Geroch, A. Held, R. Penrose, A space-time calculus 
         based on pairs of null directions, J. Math. Phys. {\bf 14} 
         874--881 (1973)

\bibitem{FSz}
         J. Frauendiener, L. B. Szabados, The kernel of the edth 
          operators on higher-genus spacelike 2-surfaces, Class. 
          Quantum Grav. {\bf 18} 1003--1014 (2001), gr-qc/0010089

\bibitem{HT} 
          G. Horowitz, K. P. Tod, A relation between local and 
          total energy in general relativity, Commun. Math. Phys. 
          {\bf 85} 429--447 (1982)

\bibitem{Sz94b} 
          L. B. Szabados, Two dimensional Sen connections and 
          quasi-local energy-momentum, Class. Quantum Grav. {\bf 11} 
          1847--1866 (1994), gr-qc/9402005

\bibitem{Wi} 
          E. Witten, A new proof of the positive energy theorem, 
          Commun. Math. Phys. {\bf 30} 381--402 (1981)

\bibitem{Ne} 
          J. M. Nester, A new gravitational energy expression and 
          with a simple positivity proof, Phys. Lett. A, {\bf 83} 
          241--242 (1981)

\bibitem{NT}
          E. T. Newman, K. P. Tod, Asymptotically flat spacetimes, in 
           {\it General Relativity and Gravitation}, Ed. A. Held, Vol 2, 
           pp 1--36, Plenum Press, New York 1980

\bibitem{Sh}
          W. T. Shaw, The asymptopia of quasi-local mass and momentum: 
           I. General formalism and stationary spacetimes, Class. 
           Quantum Grav. {\bf 3} 1069--1104 (1986) 

\bibitem{St} 
          J. Stewart, {\it Advanced General Relativity}, Cambridge 
           University Press, Cambridge 1990

\bibitem{AsKr} 
          A. Ashtekar, B. Krishnan, Dynamical horizons and their 
          properties, Phys. Rev. D {\bf 68} 104030--1-25 (2003), 
          gr-qc/0308033


\bibitem{BoFa} 
          I. Booth, S. Fairhurst, Horizon energy and angular momentum 
          from a Hamiltonian perspective, Class. Quantum Grav. {\bf 22} 
          4515--4550 (2005), gr-qc/0505049 

\bibitem{Gour} 
          E. Gourgoulhon, Generalized Damour--Navier--Stokes equations 
          applied to trapping horizons, Phys. Rev. D {\bf 72} 
          104007--1-16 (2005), gr-qc/0508003

\bibitem{Br} 
          B. D. Bramson, Relativistic angular momentum for 
           asymptotically flat Einstein--Maxwell manifolds, Proc. Roy. 
           Soc. Lond. A, {\bf 341} 463--490 (1975)

\bibitem{NP} E. T. Newman, R. Penrose, New conservation laws for zero 
           rest-mass fields in asymptotically flat spacetimes, Proc. 
           Roy. Soc. A, {\bf 305} 175--204 (1968)

\bibitem{Br78} 
          B. D. Bramson, The invariance of spin, Proc. Roy. Soc. Lond. 
          A, {\bf 364} 383--392 (1978)

\bibitem{CT} 
         P. T. Chru\'sciel, K. P. Tod, An angular momentum bound at 
         null infinity, arXiv:0706.4057

\bibitem{Da}
         S. Dain, The inequality between mass and angular momentum 
         for axially symmetric black holes, arXiv:0707.3118

\bibitem{HF}
         H. Friedrich, Gravitational fields near space-like and null 
         infinity, J. Geom. Phys. {\bf 24} 83--163 (1998)


\end{thebibliography}
\end{document}